\documentclass[aps,prx,reprint]{revtex4-2}
\usepackage{amsmath,braket,amsthm,amssymb}
\usepackage{graphicx}
\graphicspath{{figs/}}
\usepackage[hidelinks, colorlinks=true,
            urlcolor=blue, citecolor=blue]{hyperref}
\usepackage{cleveref}
\usepackage{xcolor}
\usepackage{comment}
\usepackage[toc,page]{appendix}
\usepackage{microtype}
\usepackage{thmtools}
\usepackage{thm-restate}
\usepackage{caption}
\usepackage[labelformat=simple,skip=10pt]{subcaption}
\usepackage{enumitem}
\usepackage{microtype}
\usepackage{ulem}
\usepackage[justification=raggedright,singlelinecheck=false]{caption}

\newcommand{\R}{\mathbb{R}}

\newcommand{\mix}{\text{mix}}
\newcommand{\CC}{\mathcal{C}}

\renewcommand\footnotemark{}

\newtheorem{theorem}{Theorem}

\newtheorem{lemma}[theorem]{Lemma}

\newcommand{\tr}{\mathrm{Tr}}

\newcommand{\jnote}[1]{{\color{red} {\bf Jun}: {#1}}}
\newcommand{\cnote}[1]{{\color{blue} {\bf Chaithanya}: {#1}}}

\renewcommand{\jnote}[1]{}
\renewcommand{\cnote}[1]{}

\begin{document}

\title{Fast mixing of operator-loop path-integral quantum Monte Carlo \\for stoquastic XY Hamiltonians}

\author{Chaithanya Rayudu}
\email{chaithanyarss@unm.edu}
\affiliation{Department of Physics and Astronomy, Center for Quantum Information and Control,\\ University of New Mexico, Albuquerque, New Mexico, USA}
\author{Jun Takahashi}
\email{juntakahashi@issp.u-tokyo.ac.jp}
\affiliation{The Institute for Solid State Physics, The University of Tokyo, Kashiwa, Chiba, Japan}

\begin{abstract}
Quantum Monte Carlo method with operator-loop update is a powerful technique that has been extensively used with great success in condensed matter physics.
It enables one to sample from thermal and ground states of local Hamiltonians of various spin, bosonic and fermionic systems as long as the Hamiltonian does not have a negative-sign problem.
Despite the practical success of this method, theoretical understanding of the efficiency of the algorithm has been lacking.
The operator-loop update is commonly used for path-integral formulation (Suzuki-Trotter/world-lines) of the partition function.
In this work we consider this method applied to the stoquastic (sign-problem free) XY model and prove that the mixing time of the Markov chain is polynomial in the system size and the inverse temperature.
Using the fast mixing Markov chain, we can estimate the partition functions of the Hamiltonians that we consider in a polynomial time, significantly improving upon the best known previous algorithm by Bravyi and Gosset \cite{bravyi2017polynomial}.
Our algorithm also allows for natural extensions to a wide class of empirically fast-mixing Hamiltonians.
\end{abstract}

\maketitle

\section{Introduction}
Quantum Monte Carlo (QMC) methods are a powerful class of classical algorithms in condensed matter physics, enabling unbiased simulations of large quantum systems at both zero and finite temperatures. In particular, operator-loop QMC algorithms have been used with great success to study critical phenomena and ground state properties \cite{sandvik2010computational}. Despite this empirical success, a rigorous theoretical understanding of why these algorithms converge efficiently, i.e. in time proportional to a polynomial in system size $n$ and inverse temperature $\beta$, has been lacking.

In contrast, other computational approaches, such as tensor networks, are supported by a well-developed theoretical framework. 
Their limitations can often be characterized in terms of entanglement entropy, providing clear guidance on when and why the algorithms can be expected to perform well. 
For QMC, two fundamental factors govern the feasibility of simulations: the infamous sign problem, which has been studied extensively, and the relaxation time of QMC as a Markov chain. 
While the former has received recent attention \cite{marvian2019computational, hangleiter2020easing}, rigorous results on the latter remain scarce with a few notable examples of transverse-field Ising model \cite{crosson2021rapid,bravyi15MonteCarlo}, small inverse temperature \cite{Crosson2025classicalsimulation}, restricted class of Heisenberg Hamiltonians \cite{takahashi2024rapidly}. 

A notable theoretical result in this regard is the work of Bravyi and Gosset \cite{bravyi2017polynomial}, which proves that a randomized classical algorithm can estimate the partition function of the stoquastic XY Hamiltonians in polynomial time. Their approach uses a Suzuki-Trotter decomposition very similar to the path-integral QMC, but relies on a mapping to the counting problem of weighted perfect matchings for which Jerrum and Sinclair \cite{jerrum1989approximating} gave an efficient algorithm. Despite this breakthrough, the algorithm of Jerrum and Sinclair is far from useful in practice to estimate the partition function. It also seems difficult to extend the Bravyi-Gosset algorithm to more elaborate models like the stoquastic Heisenberg model which has been raised explicitly as an open problem in \cite{gharibian7faces}.

In this work, we consider operator-loop QMC for the family of stoquastic XY Hamiltonians that Bravyi and Gosset consider in \cite{bravyi2017polynomial}. Operator-loop QMC is closer to QMC algorithms that are used in practice for a broad class of Hamiltonians \cite{sandvik2010computational} including the class of Hamiltonians we consider in this work. 
We show that the operator-loop update Markov chain mixes in polynomial time for the family of stoquastic XY Hamiltonians that we consider.
Using this, we give an upper bound of $\tilde{\mathcal{O}}(n^{22}\beta^{11}\epsilon^{-6})$ on the runtime of the algorithm to estimate the partition function, a significant improvement over $\tilde{\mathcal{O}}(n^{92}\beta^{46}\epsilon^{-25})$ runtime upper bound of the best know previous algorithm by Bravyi and Gosset \cite{bravyi2017polynomial, childs2021theory}. The technical contribution of our work is the mixing time analysis of the operator-loop update Markov chain for the stoquastic XY Hamiltonians. Our main tool for the analysis is the canonical path and multicommodity flow methods invented by Jerrum and Sinclair \cite{jerrum1989approximating, jerrum1993polynomial} to analyze the mixing time of Markov chains. 

Our work also opens up a interesting future direction to extend the efficiency results of operator-loop QMC to other systems such as the stoquastic Heisenberg Hamiltonians, Bose-Hubbard models and quantum rotors for which operator-loop QMC is empirically known to mix fast. Beyond providing algorithmic insights, our results carry physical implications. Polynomial mixing times imply rigorous bounds on correlation lengths in the QMC configurations, which in turn lead to bounds on the dynamical critical exponent $z$ for any critical points that possibly arise in the models. 
Polynomial mixing time also guarantees that the states cannot exhibit a discrete symmetry-breaking phase such as valence-bond solid (VBS) phase, since they must have exponential mixing time. 
We will discuss such implications in \cref{sec:discussion} of this paper. 

The rest of this paper is organized as follows. 
In the following subsections of the introduction, we give a semi-technical overview of the paper. 
Namely, we briefly describe what QMC is, concisely state our results, intuitively explain how the proof techniques work, and finally make direct connections with other works. 
In \cref{sec:QuantumClassicalMapping}, we explain in detail how we arrive at the semiclassical state space of the operator-loop update Markov chain starting from our target $XY$ Hamiltonian, and the relation between the problem of estimating the partition function and the problem of sampling from the semiclassical state space.
In \cref{sec:operator_update_markov_chain}, we explain in detail what the operator-loop update Markov chain is on top of the semiclassical state space. 
The sections \ref{sec:QuantumClassicalMapping} and \ref{sec:operator_update_markov_chain} together frame the QMC algorithm we will be analyzing --- the operator-loop update for the Suzuki-Trotter based path-integral QMC.
In \Cref{sec:MixingTime}, we give our main proof of the mixing time bound of the operator-loop update Markov chain where we explaining our multicommodity flow construction and its analysis.
Finally, in \cref{sec:discussion}, we discuss the connection between our result and condensed matter physics as well as other related quantum many-body research. We also list interesting open problems.

\subsection{Quantum Monte Carlo}

QMC algorithms can be separated into two conceptual building blocks. 
First, the quantum partition function $Z=\tr[\exp({-\beta H})]$ is written approximately as a summation of positive weights $W(\gamma)$ over a potentially exponentially big classical configuration space $\{\gamma\}$. 
This is called the quantum-classical correspondence, since the latter weights can be interpreted as the Boltzmann factor $\exp({-\beta E(\gamma)})$ of a classical energy function $E(\gamma)$. 
Second, a Markov chain (random walk) among the configurations $\gamma$ which can be represented as a transition probability matrix $P$. 
The Perron-Frobenius theorem guarantees that the Markov chain converges to the distribution corresponding to the targeted weights $\pi(\gamma)\propto W(\gamma)$ under mild assumptions.  All QMC algorithms are essentially built upon these two steps, and therefore has two major elements that directly affect their efficiency. 

First is whether the quantum-to-classical mapping even exists. 
There are standard procedures such as the Suzuki-Trotter decomposition and Taylor expansion, which roughly correspond to path-integral formalism and Stochastic Series Expansion (SSE) formalism respectively. 
However, it is also well known that these standard procedures do not necessarily give all non-negative weights $W(\gamma)$ depending on the Hamiltonian. This is known as the negative-sign problem \cite{sandvik2010computational}. 
{\it Stoquastic} Hamiltonians are defined as Hamiltonians with no positive off-diagonal elements in the computational basis. This guarantees that all the weights are nonnegative after the standard mappings \cite{bravyi2006stoq}. 
In this work we consider a class of XY Hamiltonians where stoquasticity is automatically satisfied by the constraints on the coefficients (see \cref{eq:stoq_XY_hamiltonian}). 

The second element that crucially affects the efficiency of QMC is the {\it mixing time} of the Markov chain which is roughly the amount of time we need to perform the random walk before we are guaranteed to be close to the equilibrium, i.e., the target distribution $\pi(\gamma) \propto \exp({-\beta E(\gamma)})$. 
The task of upper bounding the mixing time is essentially equivalent to lower bounding the spectral gap between the largest and second largest eigenvalues of the Markov chain's transition matrix $P$.
Such lower bounds are generally hard to obtain, as exemplified in the long-standing open problem of a rigorous proof for the Haldane gap. 
The problem of upper bounding the mixing time is the one we tackle in this work.

\subsection{Overview}

In our work, the state (configuration) space of QMC is defined by the quantum-to-classical mapping based on Suzuki-Trotter decomposition (defined in \cref{sec:QuantumClassicalMapping}), and we then use the ``operator-loop update'' Markov chain (also referred to as the worm update/algorithm)(defined in \cref{sec:operator_update_markov_chain}) to do a random walk over the configuration space. The operator-loop update encompasses a broad class of random walks where the general strategy employed by this family is to extend the configuration space to include configurations with (usually two) ``defects" and construct a valid Markov chain over the extended space, and then ensure that the original configuration space is not too rare to occur in the extended space. 

While this method has provided numerous success in simulating quantum thermal states, for a long time operator-loop updates were more of a practically powerful tool rather than an interest of theoretical analysis.
We believe, part of the reason came from the fact that the Markov chain is objectively more complicated than that of the classical Ising model, where a single step in the Markov chain was just to flip a single spin. 
The operator-loop update is usually regarded as a {\it nonlocal} update method, which means that it operates by creating a ``cluster'' on the configuration stochastically and flipping the entire cluster. 
The way the cluster is stochastically constructed is topological and loop-like, hence the name. 
Also the cluster in principle may grow indefinitely, which is the reason why it is regarded as a nonlocal update method.
In our analysis, we slightly modify the Markov chain from the most standard implementation of operator-loops which usually breaks detailed balance, and introduce a way to regard all intermediate steps of the cluster construction to be part of the extended configuration space random walk (see \cref{sec:operator_update_markov_chain} for details of the Markov chain). 
With this change of perspective, all steps in the Markov chain can be understood locally and becomes more amenable to analysis.

\subsubsection{Results}
We show that the mixing time of the operator-loop update Markov chain $\mathcal{M}$ (defined in \cref{sec:operator_update_markov_chain}) for a class of stoquastic XY Hamiltonians (defined in \cref{eq:stoq_XY_hamiltonian}) is upper bounded as
\begin{align}
    \tilde{\mathcal{O}}(n^{16}\beta^{8}\epsilon^{-4})
\end{align}
where $n$ is the number of qubits, $\beta$ is the inverse temperature and $\epsilon$ is the error due to the Suzuki-Trotter approximation of the partition function. Using a counting-to-sampling reduction argument (explained in \cref{sec:counting_to_sampling_reduction}) and the Markov chain $\mathcal{M}$, we can estimate the partition function $Z = \tr[\exp(-\beta H)]$ to within a $\epsilon$ precision in time
\begin{align}
\tilde{\mathcal{O}}(n^{22}\beta^{11}\epsilon^{-6})
\end{align}
The best known previous algorithm for estimating the partition function for the class of Hamiltonians that we consider is by Bravyi and Gosset \cite{bravyi2017polynomial, childs2021theory} with a runtime upper bound of $\tilde{\mathcal{O}}(n^{92}\beta^{46}\epsilon^{-25})$.

\subsubsection{Overview of techniques}
A significant portion of our technical work is bounding the mixing time of the operator-loop update Markov chain $\mathcal{M}$. 
There are only a few methods for rigorously upper bounding the mixing time of a discrete Markov chain. 
In this work, we use a technique called the {\it multicommodity flow} invented by Jerrum and Sinclair in \cite{jerrum1989approximating, sinclair1992improved} to bound the mixing time of discrete Markov chains. 
Any discrete Markov chain can be viewed as a random walk on a potentially exponentially large discrete state space. 
We can visualize this as a graph where vertices are states and directed weighted edges correspond to the nonzero transition probabilities in the Markov chain. 
The high-level idea of the {\it multicommodity flow} technique is to explicitly show that the graph is well connected and the probability masses can flow through this graph without too much congestion at any edge.
In other words, we show that probability masses can spread through the entire state space without getting stuck in a bottleneck by explicitly constructing paths/flows that probability masses can move through. 
This constructed flow is not necessarily the actual way probability spreads, and are in general artificial. The sole purpose of this construction is to show the {\it existence} of such paths, so that it leads to an upper bound on the mixing time. 

Specifically, the multicommodity flow technique requires constructing a flow of probability masses between all possible pair of states in the configuration space, say $\gamma$ and $\xi$. 
The amount of probability (mass) the flow needs to carry is the product of the stationary probabilities $\pi(\gamma)\pi(\xi)$. 
The intuition is that we want to show that a big amount of probability can be carried between configurations that have relatively high probability in the stationary distribution $\pi$, while not creating too much congestion at the small probability configurations.

Technically, flows are defined as a family of positively weighted paths for each fixed pair of configurations $\gamma,\xi$. 
For example, we can allot probability mass of 0.2 to take the path $\gamma\to\omega\to\xi$ and another 0.05 to take the path $\gamma\to\omega^{\prime}\to\xi$. 
This would be a valid flow between $\gamma$ and $\xi$ if $\pi(\gamma)\pi(\xi)=0.25$. 
Once these flows for all possible ordered pairs $(\gamma,\xi)$ are defined, 
we analyze how congested each edge is by calculating the ratio between the total flow using the edge and its {\it capacity}. 
When we have a Markov chain with transition probability $P(\gamma\to \xi)\geq0$, the capacity of that edge $\gamma\to \xi$ is $\pi(\gamma)P(\gamma\to\xi)$, which is the edge's natural amount of carrying probability, and hence called the capacity.
If the constructed multicommodity flow do not exceed this capacity by more than a polynomial factor, it intuitively implies that the natural capacity of the edges of the state space graph is enough to handle mixing i.e., connecting all states with weighted importance proportional to their equilibrium distribution. 
More rigorously, if we can upper bound the {\it congestion} of all edges with a polynomial factor, that implies a mixing time upper bound of roughly the same polynomial degree.
The congestion is defined as the sum of all the probabilities that the paths that go through a single edge $\gamma\to\xi$ are carrying, divided by the capacity of that edge (see \cref{sec:MixingTime} for details).

The multicommodity flow we construct for upper bounding the mixing time of operator-loop update Markov chain is slightly technical, yet conceptually not too complicated. 
The main observation is the topological fact that the difference between any two configurations (say, $\gamma$ and $\xi$) can be decomposed into a combination of loops and open strings.
This then can be easily turned into canonical paths through which we can route the probability $\pi(\gamma)\pi(\xi)$: we define a canonical ordering for such combination of loops and open strings, and unwind them one by one in the canonical order. Our multicommodity flow construction is only slightly more complicated than the description we give here and involves a distribution of paths to route the probability between certain pairs of configurations rather than a single canonical path.  

To complete the proof, we need to actually show that the above multicommodity flow has a polynomially bounded congestion. 
In order to do this, we rely on the idea of {\it encoding functions} introduced in \cite{jerrum1989approximating} and used in several other works e.g. \cite{jerrum1993polynomial, jerrum2004polynomial, huang2016canonical, crosson2021rapid, takahashi2024rapidly}. 
The definition of congestion (see \cref{defeq:congestion}) contains the capacity of an edge in the denominator and takes the double summation for state pairs $\gamma$ and $\xi$ in the numerator, which is the main source of potentially exponentially large value. 
The main idea behind the encoding function is to make use of a one-to-one function $\eta_{(\phi,\psi)}:(\gamma,\xi)\mapsto\zeta$ which maps from the pair of starting $\gamma$ and ending $\xi$ configurations whose canonical path passes through the particular transition edge $P(\phi\to\psi)$ to a single configuration $\zeta$. 
Thanks to the one-to-one property, we can rewrite the double summation over $\gamma$ and $\xi$ to a single summation over $\zeta$, which now has hope for comparing with the exponentially small denominator so that we arrive to a polynomial upper bound. 
From this logic, it is clear that the same strategy with some variation will still successfully prove polynomial mixing, 
e.g., allowing the function $\eta$ to be $C$-to-one mapping with $C$ only polynomially large. 
In our case, the mapping has exponentially large $C$, but we can upper bound the summed weights to only add up to a constant, thus effectively being constant-to-one mapping.

 Together, our technical contributions can be summarized as follows. 
\begin{enumerate}[leftmargin=*]
    \item Reformulating the practical operator-loop update to a detail-balance satisfying Markov chain, which allows for a local interpretation, allowing for multicommodity flow analysis. In order to do this, we extend the state space of the Markov chain to include configurations with ``defects''. 
    \item Constructing the multicommodity flow for the Markov chain we define. This relies on the observation that all differences between states in the original space can be decomposed into loops and open strings. 
    \item Obtaining a polynomial upper bound on the congestion for the multicommodity flow we construct with respect to the Markov chain. 
\end{enumerate}
By going through all of these steps rigorously, we finally arrive at an upper bound of $\mathcal O(n^{16} \beta^{8}\epsilon^{-4})$ on the mixing time of the Markov chain and an upper bound of $\mathcal O(n^{22} \beta^{11}\epsilon^{-6})$ for the algorithm to estimate the partition function of a Hamiltonian of the form in \cref{eq:stoq_XY_hamiltonian}.

\subsection{Our contribution and comparison to related research}

Although no mathematical bounding of the mixing time for the operator-loop QMC we consider in this paper has been conducted before, there are a number of related results regarding the mixing time of QMC in the broader sense.

The family of stoquastic XY Hamiltonians that we consider in this work is exactly same as the one considered in \cite{bravyi2017polynomial}, and in that sense our work can be regarded as a significant improvement over their work giving an improved algorithm with an upper bound of $\mathcal{O}(n^{22}\beta^{11}\epsilon^{-6})$ on the runtime compared to $\tilde{\mathcal{O}}(n^{92}\beta^{46}\epsilon^{-25})$ \cite{bravyi2017polynomial, childs2021theory}.  
We claim that this improvement is not merely a marginal refinement in numbers.
Most crucially, the work of \cite{bravyi2017polynomial} relied on the mapping of the partition function estimation problem to the approximate counting of weighted perfect matching. 
The analysis then uses the counting perfect matchings algorithm presented in \cite{jerrum1989approximating} as a black box, and is in that sense entirely focuses on the fact that a mapping to counting perfect matchings is possible. 
Although this connection between the XY model partition function and counting perfect matching is interesting in its own right
(especially in connection with matchgates \cite{jozsa2008matchgates,zhao2024learning}, which allows efficient classical simulation), 
the fundamental reason why such connection exists has remained a mystery for years, and because of that, extensions of their results to a broader class of Hamiltonians have been almost nonexistent. 

Interestingly, Jerrum and Sinclair's algorithm for perfect matching estimation \cite{jerrum1989approximating} is also a Markov chain, and they use their multicommodity-flow proof technique to bound the mixing time. 
Similar to the path-integral QMC configurations, perfect matchings have hard constraints which makes naive sampling very difficult. 
Again, very similar to the operator-loop update, Jerrum-Sinclair Markov chain cleverly extends the state space to allow {\it near} perfect matchings, which corresponds to adding defects. 
In some sense, the Markov chain we construct and analyze in this work can be thought of as an intermediate point between the Jerrum-Sinclair chain and the most practical directed loop update QMC \cite{syljuaasen2002quantum}. 
Naturally, focusing on specific forms of the partition function derived from the quantum-classical mapping allows for a more efficient Markov chain that is specialized. 
However, most crucially, operator-loop update Markov chains are in general not a simple application of perfect matching sampling, but rather the other way around: they are empirically known to perform efficiently for a far broader family of Hamiltonians. 
This potential extendability is pronounced when we consider extending the fast mixing result of this work and \cite{bravyi2017polynomial} to include isotropic Heisenberg models. 
On the one hand, operator-loop updates are known to perform just as well for stoquastic Heisenberg models \cite{syljuaasen2002quantum}. 
On the other hand, {\it if} the construction of \cite{bravyi2017polynomial} allows for $-ZZ$ interactions, it should allow for $+ZZ$ interactions as well, 
which then should not be fast-mixing since the ground state problem will be $\sf{StoqMA}$-hard \cite{10.5555/3179553.3179559}. 
By not relying on the perfect matching mapping, we were able to have a polynomial run time proof that is potentially far more extendable, has considerably better polynomial degree, and more direct connections with computational condensed matter physics. 

The aforementioned extendability to Heisenberg models becomes even more interesting when we consider the open problem addressed in \cite{gharibian7faces}: the computational complexity of the ground state problem for bipartite quantum max cut, which is essentially the same as stoquastic Heisenberg models (see \cref{sec:HeisenbergOpenProblem} for more details). 
In this direction, there has been some recent effort to show fast-mixing of a QMC algorithm specialized for the stoquastic isotropic Heisenberg model \cite{takahashi2024rapidly} where the authors were only able to prove fast mixing for graphs with one side of the bipartition as $\mathcal O (\log n)$. 
While the successful application of the canonical path method by Jerrum and Sinclair to a practical QMC method, and a super constant bipartition size was novel and nontrivial, unfortunately the restriction to $\mathcal O(\log n)$ sized sublattices makes the result not directly relevant to condensed matter physics models. 
On the contrary, the results in this work applies to all interaction graphs, including very commonly considered square lattices, triangular lattices, etc. 

Turning our attention to other QMC fast-mixing proofs, there has been a few results \cite{crosson2021rapid,Crosson2025classicalsimulation} on the fast mixing of QMC for stoquastic Hamiltonians restricted to 1-dimension or high temperature.
If we set the coefficients of our Hamiltonian to $b=0$ (see \cref{eq:stoq_XY_hamiltonian}), it is clear that our setup can handle transverse field Ising models (TFIM) too (but with a basis rotation of $X\leftrightarrow Z$) as long as the TFIM Hamiltonian does not contain longitudinal fields. 
While both works include longitudinal fields for the TFIM and therefore prohibits truly direct comparison, \cite{crosson2021rapid} is restricted to 1d geometry and relatively high inverse temperature $\beta\sim\mathcal O(\log n)$ and \cite{Crosson2025classicalsimulation} where $\beta\sim\mathcal O(1)$ is required. 
The fact that they only consider local update QMC heavily restricts what is even provable. Namely, path-integral QMC with only simple single-flip updates (no operator-loop updates) are empirically known to ``freeze'' in the classical limit with transverse field $\Gamma\to0$. Therefore, a general fast mixing proof must be impossible and therefore is a fundamental limitation of these methods. 
On the other hand, since operator-loop update methods are one of the currently known state of the art QMC frameworks and does not suffer from such freezings, we believe our results ultimately should be able to be extended to TFIM with longitudinal fields, or equivalently, stoquastic transverse fields in \cref{eq:stoq_XY_hamiltonian}. This is left as an interesting open problem. 

Finally, we note that fast-mixing results are also known for the classical 8-vertex and related models \cite{cai2021fptas,cai2019sixvertex} via variants of similar Jerrum-Sinclair chain. 
Although in some limiting parameter regime the 1-dimensional $XY$ chain partition function is exactly mappable to that of the 8-vertex model of a 2-dimensional square lattice, this correspondence is restricted to special geometries and does not extend to general settings.
In contrast to the inherently classical nature of the 6- and 8-vertex models, the $XY$ model is intrinsically quantum, and admits a wider range of natural extensions of QMC that remain amenable to fast-mixing analysis.

\section{Approximation of partition function}
\label{sec:QuantumClassicalMapping}

In this work we consider the problem of computing the partition function $Z := \tr[e^{-\beta H}]$ of an $n$-qubit stoquastic XY Hamiltonian of the following form
\begin{align} \label{eq:stoq_XY_hamiltonian}
    H = \sum_{1\leq i<j\leq n}  (-a_{ij} X_iX_j +b_{ij}Y_iY_j) + \sum_{i=1}^{n} d_iZ_i
\end{align}
where $ |b_{ij}| \leq a_{ij} \leq 1/2$ and $|d_i|\leq 1$. Without loss of generality, we will assume that the inverse temperature $\beta \geq 1$ since we can rescale the Hamiltonian otherwise.

\subsection{Path integral formulation}
\label{sec:PI_def}
In the path-integral formulation, we begin by approximating the partition function using the Suzuki-Trotter approximation. The following theorem, proved in \cite{bravyi2017polynomial,childs2021theory}, gives a bound on the error due to the approximation in terms of the number of Trotter steps in the Suzuki-Totter approximation.
\begin{theorem}[\cite{bravyi2017polynomial,childs2021theory}]
    \label{thm:trotterization}
    Given a Hamiltonian $H$ of the form in \cref{eq:stoq_XY_hamiltonian} and a positive real number $\epsilon$, we can choose an integer $L = \mathcal{O}(n^2\beta^2\epsilon^{-1})$ and approximate the partition function $Z(\beta, H)$ with the Trotterized partition function $\tilde{Z}(\beta, H)$, defined as
    \begin{gather}
        \tilde{Z}(\beta, H) := \tr[(C C^\dagger)^L] \label{eq:trotterized_partition_function}
    \end{gather}
    where
    \begin{align}
        C = \prod_{1\leq i  \leq j \leq n} \left(I+\delta \left(a_{ij} X_iX_j -b_{ij}Y_iY_j\right)\right) \nonumber\\
        \times \prod_{1 \leq i \leq n} \left(I- \delta\, d_i Z_i\right) \label{eq:trotter_expansion}
    \end{align}
    and $\delta = \beta/2L$, such that
    \begin{align}
        e^{-\epsilon/4} Z \leq \tilde{Z} \leq e^{\epsilon/4} Z.
    \end{align}
\end{theorem}
In \cite{bravyi2017polynomial}, the operators $I+\delta \left(a_{ij} X_iX_j -b_{ij}Y_iY_j\right)$ are approximated by matchgates. Here we do not need that approximation but instead approximate the operators only to first order in $\delta$. Additionally, we don't need the strict guarantees on the partial operator sequences as is required in \cite[lemma~1]{bravyi2017polynomial}. 
The proof of \cref{thm:trotterization} is implicit from the proofs of corresponding approximation theorems and lemmas in \cite{bravyi2017polynomial, childs2021theory}.

Using \cref{eq:trotter_expansion}, let us rewrite $(CC^\dagger)^{L}$ as 
\begin{align}
    \label{eq:PI_string_of_operators}
    (CC^\dagger)^{L} = \prod_{m=1}^{M} O_m
\end{align}
where $O_m$ are one or two qubit operators from the product in \cref{eq:trotter_expansion}. Let $M_1$ be the number of one qubit operators and $M_2$ be the number of two qubit operators in the product $\prod_{m=1}^{M} O_m$ and let $M = M_1 + M_2 = \mathcal{O}(n^4\beta^{2}\epsilon^{-1})$ be the total number of operators. Substituting \cref{eq:PI_string_of_operators} back into \cref{eq:trotterized_partition_function} and inserting resolutions of identities between the sequence of operators $\{O_m\}_{m=1}^M$, we can rewrite the Trotterized partition function
\begin{align}
    \tilde{Z} &= \sum_{\sigma_1 \in \{0,1\}^n} \bra{\sigma_1} (CC^\dagger)^L\ket{\sigma_1}\\
    &= \sum_{\substack{\sigma_1,\sigma_2,...,\sigma_M\\ \sigma_{M+1} := \sigma_{1}}}\quad \prod_{m = 1}^{M} \bra{\sigma_m}O_m \ket{\sigma_{m+1}}, \quad \\
    &= \sum_{\sigma_1,\sigma_2,...,\sigma_M} W(\sigma_1,...,\sigma_M)
\end{align}
as a summation of weights
\begin{gather}
    W(\sigma_1,...,\sigma_M) := \prod_{m = 1}^{M} \bra{\sigma_m}O_m \ket{\sigma_{m+1}} \label{eqdef:weight_of_config_PI}
\end{gather}
on strings of computational basis states $(\sigma_1,...,\sigma_M)$. The extra dimension induced by the string of states $(\sigma_1,...,\sigma_M)$ is usually referred to as the imaginary time direction.

We will refer to a single instance of a string of computational basis states $(\sigma_1,...,\sigma_M)$ as a `path-integral configuration'. Given that the Hamiltonian $H$ in \cref{eq:stoq_XY_hamiltonian} is stoquastic, the weights $W$ are non-negative for all configurations. Using these non-negative weights, let us define a probability distribution $\pi_{0}$ over the path-integral configuration space as
\begin{align}
    \label{eq:distribution_pi0_PI}
    \pi_{0}(\sigma_1,...,\sigma_M) := \frac{W(\sigma_1,...,\sigma_M)}{\tilde{Z}}.
\end{align}
Using the counting-to-sampling reduction, as described in \cref{sec:counting_to_sampling_reduction}, we can reduce the problem of estimating the Trotterized partition function $\tilde{Z}$ to sampling from the distribution $\pi_{0}$ at different inverse temperatures $\beta$.

\subsection{Diagrammatic notation}
\label{sec:diagrammatic_notation}
\begin{figure}[h]
    \centering
    \includegraphics[scale=0.75]{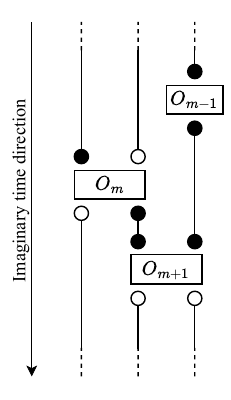}
    \caption{Diagrammatic representation of a slice in the imaginary time direction of a configuration for 3-qubits. The empty and filled circles represent $\ket{0}$ and $\ket{1}$ states respectively. The rectangular boxes across the vertical lines represent operators. The operators shown in the figure contribute $\bra{101}O_{m-1}\ket{101}$ $\bra{101}O_m\ket{011}$ $\bra{011}O_{m+1}\ket{000}$ as a factor towards the weight of the configuration.}
    \label{fig:valid_config_example}
\end{figure}

In the rest of this paper, we use a diagrammatic notation when it is convenient to represent a configuration of a string of states $\{\sigma_1,...\sigma_M\}$ and the operators $\{O_1,...,O_M\}$ interlaced between the states. \Cref{fig:valid_config_example} is an example of one such diagram. The vertical lines in the figure represent qubits with the imaginary time direction going from top to bottom. The empty and filled circles represent $\ket{0}$ and $\ket{1}$ states of the corresponding qubits at that imaginary time slice. The rectangular boxes represent either two qubit or one qubit operators depending on whether the box has four or two `legs' connected to it. Whenever two circles are connected with a vertical line, they both have to be either empty or filled to maintain consistency in the imaginary time direction. In the path-integral formulation, we assign a weight $\bra{\sigma_m}O_m \ket{\sigma_{m+1}}$ to the box representing the operator $O_m$ and the total weight of a diagram is the product of weights of all the boxes.

We call two operators $O_k$ and $O_{l}$ `neighbors' whenever they share a qubit that they act on and no other operator in between operators $O_k$ and $O_l$ act on that shared qubit. Diagrammatically, this means one of the legs from the box representing $O_k$ is directly connected to one of the legs of the box representing $O_l$ with a vertical line. In \cref{fig:valid_config_example}, the operator $\{O_m, O_{m+1}\}$ are neighbors, and $\{O_{m-1},O_{m+1}\}$ are also neighbors. 

\begin{figure}[h]
    \centering
    \includegraphics[scale=0.75]{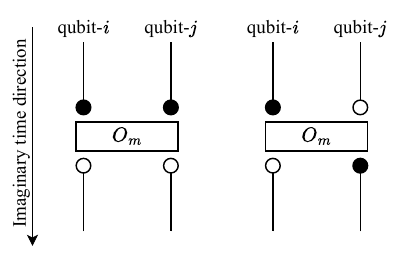}
    \caption{Diagrammatic representations of a two-qubit operator $O_m$ with weights $\bra{11}O_m\ket{00}$ and $\bra{10}O_m\ket{01}$ where $O_m = I+\delta \left(a_{ij} X_iX_j -b_{ij}Y_iY_j\right)$. The other two possible non-zero matrix elements $\bra{00}O_m\ket{11}$ and $\bra{01}O_m\ket{10}$ are obtained by replacing the empty circles with filled circles and vice versa.}
    \label{fig:off_diagonal_two_qubit_flips}
\end{figure}

A two qubit operator of the form $I+\delta \left(a_{ij} X_iX_j -b_{ij}Y_iY_j\right)$ is Hamming weight parity preserving, bit-flip symmetric and has the matrix form given in \cref{eq:two_operator_matrix_form} in the computational basis of the qubits $i$ and $j$. For this two qubit operator, there are a maximum of four possible off-diagonal elements with non-zero weights in addition to the four diagonal elements in the computational basis. Two of those four off-diagonal elements are diagrammatically represented in \cref{fig:off_diagonal_two_qubit_flips}. If $a_{ij} = b_{ij}$ or $a_{ij} = -b_{ij}$, then only one of the two will have a non-zero weight.

\section{Markov Chains}
\label{sec:operator_update_markov_chain}
In this section, we define the Markov chain that is used to sample from the distribution defined in \cref{eq:distribution_pi0_PI} of path-integral configurations.
The actual update method of the Markov chain within the configuration space is just as essential as the mapping of the quantum partition function into a classical one as we have explained in \cref{sec:QuantumClassicalMapping}.

\subsection{Markov chain for path-integral configuration space}
\label{sec:operator_update_markov_chain_PI}

\begin{figure}[h]
    \centering
    \includegraphics[scale=0.75]{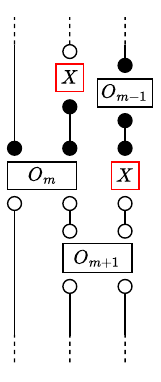}
    \caption{Diagrammatic representation of $\CC_2$ configuration with the additional Pauli-$X$ operator represented using red boxes with $X$ in them. }
    \label{fig:example_of_extended_config}
\end{figure}

Configurations in the path-integral formulation are defined by the string of states $(\sigma_1, \sigma_2...,\sigma_M)$.
We refer to the set of configurations with non-zero probability in $\pi_{0}$ as `valid configurations' and denoted the set by $\mathcal{C}_0$. To sample from the set of valid configurations $\mathcal{C}_0$ of the distribution $\pi_{0}$ using a Markov chain, we extend the configuration space to include configurations with two additional Pauli-$X$ operators inserted between the sequence of operators $\{O_{m}\}_{m=1}^{M}$ such that the two Pauli-$X$ operators are not neighbors. These additional Pauli-$X$ operators are commonly referred to as `worm heads', or `holes', and we will refer to the set of configurations with two worm heads as $\mathcal C_2$. 
\Cref{fig:example_of_extended_config} is an example of one such configuration.
We will identify the extended configurations in $\mathcal C_2$ to be the same if the inserted Pauli-$X$ operators can be moved to the same position without passing through any operators that acts on the qubit it is sitting in. 
Diagrammatically, we are taking the equivalence class of all extended operator strings $\{O_m\}^{M+2}_{m=1}$ with two additional $X$ operators to be the same if they can be made same by moving the $X$ operators along the imaginary time direction without passing through another operator. 
In other words, the worm heads should be thought of as technically not defined as an operator within the sequence but rather two points in the diagram that can move freely until it bumps into an operator.

We define the weight for an extended configurations as product of operator weights similar to the weight for a configurations in $\mathcal{C}_{0}$ in \cref{eqdef:weight_of_config_PI}. Since all the entries of a Pauli-$X$ operator in the computational basis are equal to 1, the Pauli-$X$ operators don't contribute any non-trivial factors to the total weight of a configuration.
Let $\mathcal{C}_2$ be the set of valid extended configurations with non-zero weights. Let us define a new probability distribution $\pi: \CC_0\cup\CC_2\rightarrow\R^+$ as
\begin{align}
    \label{eq:distribution_pi_PI}
    \pi(\gamma) := 
    \begin{cases}
        \frac{W(\gamma)}{W_{\text{total}}}\quad &\text{if $\gamma \in \mathcal{C}_0$}\\[1ex]
        \frac{2}{M_1 +2 M_2}\frac{W(\gamma)}{W_{\text{total}}} \quad &\text{else if $\gamma \in \mathcal{C}_2$}
    \end{cases}.
\end{align}
where
\begin{align}
W_{\text{total}} = \sum_{\gamma \in \mathcal{C}_0} W(\gamma) + \frac{2}{(M_1 +2 M_2)} \sum_{\gamma \in \mathcal{C}_2} W(\gamma).
\end{align}
The additional factor of $2/(M_1 +2 M_2)$ for the configurations in $\mathcal{C}_2$ respectively is needed to satisfy the detailed balance condition in the Markov chain that we define shortly. The distribution $\pi$ conditioned on the set of valid configurations $\mathcal{C}_0$ is equal to $\pi_{0}$ defined in \cref{eq:distribution_pi0_PI}. Therefore if we can efficiently sample from the distribution $\pi$ and the ratio of probability of configurations in $\mathcal{C}_2$ to the ratio of probability of configurations in $\mathcal{C}_0$ is polynomially bounded as shown in the following theorem then we can efficiently sample from the distribution $\pi_{0}$ with a polynomial sample overhead.
\begin{lemma}
\label{thm:polynomial_bound_on_extended_config_weight_PI}
Under the conditions of the \cref{thm:trotterization} in the path-integral formulation and with the definitions of weights of valid configurations $\CC_0$ and valid extended configurations $\CC_2$, the ratio of probability of configurations in $\CC_2$ to the probability of configurations in $\CC_0$ in the distribution $\pi$ is bounded as $\mathcal{O}(n^4\beta^{2}\epsilon^{-1})$.
\end{lemma}
The proof of \cref{thm:polynomial_bound_on_extended_config_weight_PI} is given in \cref{appendix:proof_of_polynomial_bound_on_extended_config_weight_PI}. Note that a similar result is proved in \cite{bravyi2017polynomial} and improved in \cite{childs2021theory} where the ratio of sum of weights of configurations in $\CC_2$ to sum of weights of configurations in $\CC_0$ is shown to be $\mathcal{O}(n^8\beta^4\epsilon^{-2})$. The improvement we get in \cref{thm:polynomial_bound_on_extended_config_weight_PI} is due to the additional factor of $2/(M_1 +2 M_2)$ weight for the configurations in $\CC_2$ which is absent in \cite{bravyi2017polynomial}.

Let us now define the Markov chain, denoted by $\mathcal{M}$, based on operator-loop update that has the $\pi$ as the stationary distribution. The allowed transitions in the Markov chain $\mathcal{M}$ are transitions between two configurations in $\mathcal{C}_0 \cup \mathcal{C}_2$ that only differ in a small neighborhood, and these transitions can be viewed as either inserting the two Pauli-$X$ operators between the sequence of operators $\{O_{m}\}_{m=1}^{M}$ to go from a configuration in $\mathcal{C}_0$ to another in $\mathcal{C}_2$ or moving the Pauli-$X$ operators across their neighboring operators to go from a configuration in $\mathcal{C}_2$ to another in either $\mathcal{C}_2$ or $\mathcal{C}_0$. Recall that the weight of the configurations in $\mathcal{C}_0$ and $\mathcal{C}_2$ is defined as the product of operator weights $\bra{\sigma_m}O_m \ket{\sigma_{m+1}}$. The pairs of configurations that can be transitioned between each other within one step of the random walk have a maximum of one operator weight difference. This allows us to make the transitions as heat-bath updates based on a single operator weight.

Let $\xi_t$ denote the configuration at the current time step $t$ in the random walk. The transitions of the Markov chain $\mathcal{M}$ are as follows.
\begin{enumerate}[leftmargin=*]
    \item If the current configuration $\xi_t$ belongs to $\mathcal{C}_0$ then choose one of the $4M_2+2M_1$ legs of the operators $\{O_t\}_{t=1}^{M}$ uniformly at random in $\xi_t$ and insert a Pauli-$X$ operator at that chosen leg. Say the chosen leg corresponds to operator $O_m$. Using a heat-bath update, insert a second Pauli-$X$ operator on one of the legs of the operator $O_m$ while flipping the qubits along the way. \Cref{fig:markov_chain_moves_1} and \ref{fig:markov_chain_moves_2} shows examples of this transition for a two qubit and one qubit operators. If this move is successful in changing the current configuration, then the new configuration will belong to $\mathcal{C}_2$.
    \begin{figure}[h!]
        \centering
        \begin{subfigure}[c]{0.45\textwidth}
            \centering
            \includegraphics[scale=0.75]{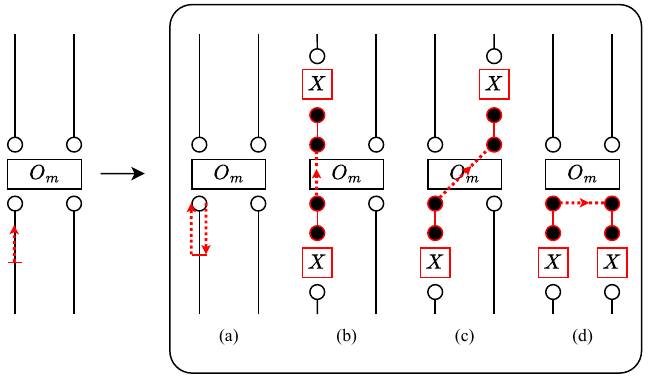}
            \caption{Example of different ways two Pauli-X operators can be inserted starting at one of the legs of a two qubit operator $O_m$ with weight $\bra{00}O_m\ket{00}$.}
            \label{fig:markov_chain_moves_1}
        \end{subfigure}
        \hspace{1cm}
        \begin{subfigure}[c]{0.45\textwidth}
            \centering
            \includegraphics[scale=0.75]{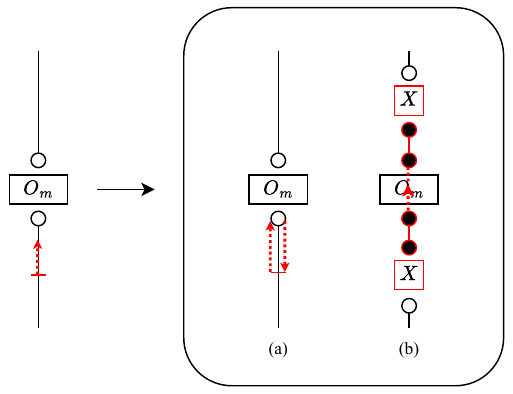}
            \caption{Example of inserting two Pauli-X operators staring at one of the legs of a one qubit operator $O_m$ with weight $\bra{0}O_m\ket{0}$.}
            \label{fig:markov_chain_moves_2}
        \end{subfigure}
        \caption{Examples of inserting two additional Pauli-$X$ operators at a two qubit and one qubit operators.}
        \label{fig:move_from_C0}
    \end{figure}

    \item If the current configuration $\xi_t$ belongs to $\mathcal{C}_2$ then choose one of the four legs of the two additional Pauli-$X$ operators uniformly at random. Using a heat-bath update, move the corresponding Pauli-$X$ operator in the direction of the chosen leg onto one of the legs of the neighboring $O_m$ operator while flipping the qubits along the way. \Cref{fig:markov_chain_moves_3} and \ref{fig:markov_chain_moves_4} shows examples of this transition for a two qubit and one qubit operators $O_m$. Whenever the two additional Pauli-$X$ operators are neighbors after this move, we remove both the operators while making the states of the qubit consistent in the imaginary time direction.
    \begin{figure}[h!]
        \centering
        \begin{subfigure}[c]{0.45\textwidth}
            \centering
            \includegraphics[scale=0.75]{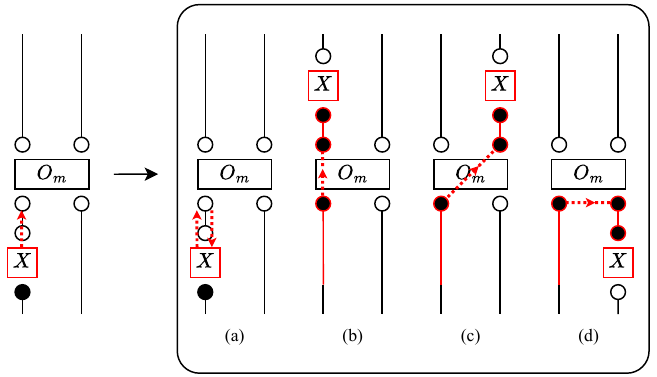}
            \caption{Example of different ways a Pauli-X operator that is present on a leg of a two qubit operator $O_m$ with weight $\bra{00}O_m\ket{00}$ can be moved onto other legs  or stay on the same leg of the operator $O_m$ while flipping the qubits along the way.}
            \label{fig:markov_chain_moves_3}
        \end{subfigure}
        \hspace{1cm}
        \begin{subfigure}[c]{0.45\textwidth}
            \centering
            \includegraphics[scale=0.75]{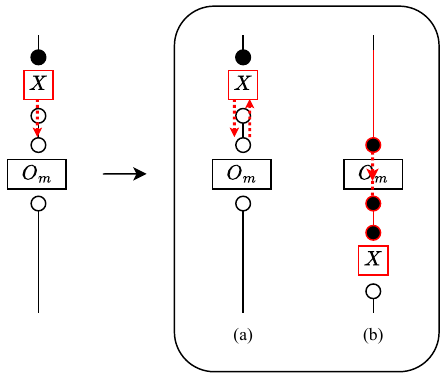}
            \caption{Example of different ways a Pauli-X operator that is present on a leg of a one qubit operator $O_m$ with weight $\bra{0}O_m\ket{0}$ can be moved onto other legs or stay on the same leg of the operator $O_m$ while flipping the qubits along the way.}
            \label{fig:markov_chain_moves_4}
        \end{subfigure}
        \caption{Examples of moving a Pauli-$X$ operator from one leg of an operator onto another leg of the operator.}
        \label{fig:move_from_C2}
    \end{figure}
\end{enumerate}
Since the above moves are made with a heat-bath update, there is a possibility of rejecting the changes and staying in the same configuration, setting $\xi_{t+1} = \xi_t$. We can verify that the stationary distribution of the above Markov chain $\mathcal{M}$ is indeed the distribution $\pi$, defined in \cref{eq:distribution_pi_PI}, by verifying that the local detailed balance condition is satisfied. This also implies that the Markov chain $\mathcal{M}$ is reversible.
We also note that the Markov chain $\mathcal{M}$ is aperiodic.

The main technical result of our work is the following bound the mixing time of the Markov chain $\mathcal{M}$.
\begin{theorem}
    \label{thm:bound_on_mixing_time}
    The mixing time of the Markov chain $\mathcal{M}$ with the initial state 
    \begin{align}
        (\sigma_1, \sigma_2...,\sigma_M) = (\ket{0}^{\otimes n}, \ket{0}^{\otimes n}, ...,\ket{0}^{\otimes n})
    \end{align} 
    is bounded as 
    \begin{align}
        c_{\min}\tilde{\mathcal{O}}(n^{16}\beta^{8}\epsilon^{-4})
    \end{align} 
    where $c_{\min} = \left(\min_{i,j}\{a_{ij}-b_{ij}, a_{ij}+b_{ij}\}\right)$.
\end{theorem}

Recall that using the counting-to-sampling reduction that we explain in \cref{sec:counting_to_sampling_reduction}, we reduced the problem of estimating the partition function to taking $\tilde{\mathcal{O}}(n^6\beta^3\epsilon^{-2})$ many samples. Since we can obtain each sample within $c_{\min}\tilde{\mathcal{O}}(n^{16}\beta^{8}\epsilon^{-4})$ time by \cref{thm:bound_on_mixing_time}, we get an upper bound of
\begin{align}
    c_{\min}\tilde{\mathcal{O}}(n^{22}\beta^{11}\epsilon^{-6})
\end{align}
on the total runtime of the algorithm to estimate the partition function $Z(\beta, H)$.

\section{Mixing time analysis}\label{sec:MixingTime}

In this section we prove that the mixing time of the Markov chain $\mathcal{M}$, defined in \cref{sec:operator_update_markov_chain}, is bounded as $\mathcal{O}(n^{16}\beta^{9}\epsilon^{-4})$. Our proof is based on the multicommodity flow technology of \cite{jerrum1989approximating, sinclair1992improved} used to bound the mixing times of Markov chains. 

\subsection{Bound on mixing time using multicommodity flow}
Let $P$ denote the transition matrix of the Markov chain $\mathcal{M}$ with a state space $\Omega$. The mixing time $t_{\text{mix}}(\delta)$ is defined as
\begin{gather}
    t_{\mix}(\delta) := \max_{x \in \Omega}\,\,t_{\mix}(x,\delta) \\
    \text{where}\quad t_{\mix}(x,\delta) := \min\{t: ||P^tx-\pi||_{\text{TV}} \leq \delta\}, 
\end{gather}
where we slightly abuse the notation to treat $x$ on the right hand side to be the Kronecker distribution for the state $x$. 
$\pi$ is the equilibrium probability distribution which the Markov chain $\mathcal{M}$ converges to, which in our case can be proven to be unique because the operator-loop update Markov chain satisfies the detailed balance condition, and it is aperiodic and ergodic. 

We can upper bound the mixing time $t_{\mix}$ using the multicommodity flow technique as follows. Let us consider the directed graph $G$ with the vertex set as $\Omega$ and edges $E := \{(z,z'): z,z'\in \Omega, P(z,z') > 0\}$ denoting the set of possible transitions of the Markov chain. For every pairs of states $x$ and $y$ in $\Omega$, we want to route $\pi(x)\pi(y)$ amount of probability mass (commodity) from $x$ to $y$ using a set of paths $\mathcal{P}_{x,y}$ that start at $x$ and end at $y$ in the graph $G$ such that
\begin{align}
    \sum_{p \in \mathcal{P}_{x,y}} f_{x,y}(p) = \pi(x)\pi(y)
\end{align}
where $f_{x,y}: \mathcal{P}_{x,y} \rightarrow \R_{\geq 0}$ is the distribution of probability mass flow. We define this weighted set of paths for all pairs $x,y\in\Omega$ and call them $\mathcal{P}=\{\mathcal{P}_{x,y}\}_{x,y}$ altogether. The congestion $q$ through an edge $(z,z') \in E_{P}$ is then defined as
\begin{align}
    \label{defeq:congestion}
    q(z,z') :=  \frac{1}{\pi(z)P(z, z')}\sum_{x,y \in \Omega}\,\, \sum_{p\,:\, (z, z')\in p \in \mathcal{P}_{xy}} f_{x,y}(p).
\end{align}
Let $q := \max_{(z,z') \in E} q(z,z')$ be the maximum congestion of any edge which is dependent on $\mathcal P$. In our case  $P(z,z')$ is lower bounded by a poly$(n,\beta,\epsilon^{-1})$ for all $P(z,z') \neq 0$. The following theorem, based on \cite{sinclair1992improved}, gives a bound on the mixing time $t_{\mix}$ in terms of the maximum congestion $q$.
In other words, the task of upper bounding the mixing time can be transformed into a task of constructing a clever (enough) $\mathcal P$.

\begin{theorem}
    \label{thm:mixing_time_theorem}
    For any lazy, ergodic Markov chain and any choice of flows, the initial state dependent mixing time $t_{\mix}(x, \delta)$ satisfies
    \begin{align}
        t_{\mix} (x, \delta) \leq ql\left(\,2\ln(\delta^{-1})+\ln\left[\pi(x)^{-1}\right]\,\right)
    \end{align}
    where $l$ is the length of the longest path among $\mathcal{P}$, and $x$ is the initial state of the Markov chain.
\end{theorem}
In our case, the operator loop Markov chain $\mathcal{M}$ (defined in \cref{sec:operator_update_markov_chain}) is not lazy but we can make it lazy and increase the mixing time by only a factor $2$.

\subsection{Congestion analysis of operator-loop update Markov chain for path-integral formalism}
\label{sec:congestion_analysis_PI}

\begin{figure}[h]
    \centering
    \includegraphics[scale=0.75]{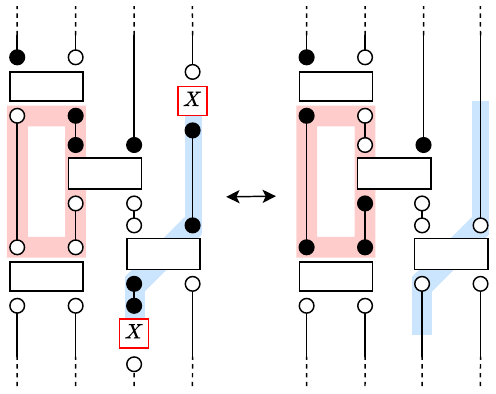}
    \caption{An example of the difference between two configurations for $\mathcal M$. The operator structure of the diagram (explained in \cref{sec:diagrammatic_notation}) is fixed, so the only differences are in the states of the qubits as highlighted. Noting the conservation of Hamming weight parity along the imaginary time for $\mathcal C_0$ configurations, one can see that the differently colored sections will always be decomposable to loops or a single string in the case of comparing with a $\mathcal C_2$ configuration.}
    \label{fig:diff_two_configs}
\end{figure}

\begin{figure}[h]
    \centering
    \includegraphics[scale=0.75]{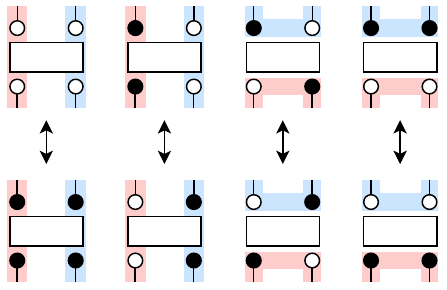}
    \caption{Vertical or horizontal decomposition into loops and paths at an operator with all four its legs different between two configurations based on the state of the qubits.}
    \label{fig:loop_decomposition}
\end{figure}

Let us make a few observations that are useful for the construction of the probability mass flow $\mathcal{P}$ for lower bounding the mixing time of the Markov chain $\mathcal{M}$. 
Given any two configurations, the difference between them can be decomposed into a set of loops and open paths where flipping the qubits on the loops and paths takes one configuration to the other as shown in the example in \cref{fig:diff_two_configs}.
The two qubit operators $O_m$ of the form $I+\delta \left(a_{ij} X_iX_j -b_{ij}Y_iY_j\right)$ will have either zero or two or four of its legs different between any two configurations with non-zero weights but never only one or three. If an operator $O_m$ has qubits on two of its legs different between two configurations, there is no ambiguity in how the loops are constructed which is the case in \cref{fig:diff_two_configs}. But if an operator has all the four legs different between two configurations, then there are multiple ways we can form the loops. When there is such ambiguity, we fix the way we decompose into loops at the operators depending on the states of the qubits on the legs of the operator as shown in \cref{fig:loop_decomposition}.

With the difference in any two configurations decomposed into loops and open paths, we start by constructing canonical paths between every two configurations $x, y \in \CC_0\cup\CC_2$ where at least one of the configurations is in $\mathcal{C}_0$, say $x \in  \mathcal{C}_0$ and $y \in \CC_0\cup\CC_2$. We use a single canonical path, denoted by $cp(x,y)$, to route the probability mass  $\pi(x)\pi(y)$ from $x$ to $y$ for these cases. We construct this canonical path starting from $x$ as a sequence of configurations obtained by first `unwinding' the loops in a canonical order and unwind the path to reach $y$. Unwinding a path or a loop consists of creating the one or two additional Pauli-$X$ operators corresponding to `worm heads' at a canonical point on the loop or the path and moving the Pauli-$X$ operators across the loop or the path while flipping the qubits along the way until the Pauli-$X$ operators meet and annihilate each other or they are annihilated at a single qubit operator $O_m$. 
The canonical path from $y$ to $x$ is obtained by unwinding in the reverse order.

For the sake of congestion analysis, along with $\CC_0$ and $\CC_2$, we will need the set of configurations with 4 Pauli-$X$ operators, denoted by $\mathcal{C}_4$, where no two Pauli-$X$ operators are neighbors. We define the weights for the configurations in $\mathcal{C}_4$ similar to weights for the configurations in $\mathcal{C}_0$ and $\mathcal{C}_2$. To estimate the congestion created by the probability flow through the canonical paths that we just defined, we use the encoding function technique to bound the flow going through any single edge. For every single edge $(z, z')$, we construct an encoding function $\eta_{(z,z')}$ that maps every canonical path $cp(x,y)$ that goes through $(z,z')$ to a configuration $\eta_{(z,z')}(x,y) \in  \mathcal{C}_0 \cup \mathcal{C}_2\cup\mathcal{C}_4$ such that
\begin{align}
    \label{eq:encoding_function_condition_PI}
    W(x)W(y) \leq \,W(z)W(\eta_{(z,z')}(x,y)).
\end{align}
The encoding function $\eta_{(z,z')}:(\CC_0\cup\CC_2)\times(\CC_0\cup\CC_2)\rightarrow\CC_0\cup\CC_2\cup\CC_4$ is {\it almost} one-to-one. We will discuss the construction of the encoding function $\eta_{(z,z')}$ and its properties in \cref{sec:encoding_function_construction}.

Let us sum over the probability mass flow through all the canonical paths that go through the edge $(z,z')$ to bound the congestion as follows
\begin{align}
   & (W_{\textup{total}})^2 \sum_{\substack{x,y\,:\, \left( x  \in\, \CC_0 \text{ or } \, y \in \CC_0\right),\\ \left(x,y \in \CC_0\cup\CC_2\right) \text{ and }(z,z') \in cp(x,y)}}\pi(x)\pi(y) \nonumber\\[4ex]
   &= \sum_{\substack{x,y\,:\, \left( x  \in\, \CC_0 \text{ and } \, y \in \CC_0\right)\\ \text{ and }(z,z') \in cp(x,y)}}W(x)W(y) \nonumber\\
   &~~~~+ \sum_{\substack{x,y\,:\, \text{exactly one of } x,y \in \CC_2,\\ \left(x,y \in \CC_0\cup\CC_2\right) \text{ and }(z,z') \in cp(x,y)}}\frac{2W(x)W(y)}{(M_1 +2 M_2)}\label{eq:congestion_analysis_bound_PI_1}\\[4ex]
   &\leq~~  \sum_{\substack{x,y\,:\, \left( x  \in\, \CC_0 \text{ and } \, y \in \CC_0\right)\\ \text{ and }(z,z') \in cp(x,y)}}W(z)W(\eta_{(z,z')}(x,y)) \nonumber\\
   &~~~~+
   \sum
   _{\substack{x,y\,:\, \text{exactly one of } x,y \in \CC_2,\\ \left(x,y \in \CC_0\cup\CC_2\right) \text{ and }(z,z') \in cp(x,y)}}
   \frac{2W(z)W(\eta_{(z,z')}(x,y))}{(M_1 +2 M_2)}\label{eq:congestion_analysis_bound_PI_2}
\end{align}
where the inequality (\ref{eq:congestion_analysis_bound_PI_2}) is obtained by substituting the inequality (\ref{eq:encoding_function_condition_PI}) into \cref{eq:congestion_analysis_bound_PI_1}.
We shall now consider two cases: either $z\in\CC_0$ or $z\in\CC_2$. 
The analysis here relies on the properties of the encoding function $\eta$ which we will be detailed in \cref{sec:encoding_function_construction}. 

\begin{enumerate}[leftmargin=*]
    \item If $z \in \CC_0$ and if $x,y\in \CC_0$, then $\eta_{(z,z')}(x,y) \in \CC_0$. But if $z \in \CC_0$ and exactly one of $x,y \in \CC_2$ while the other $\in \CC_0$ then $\eta_{(z,z')}(x,y) \in \CC_2$. Along with the fact that $\eta_{(z,z')}$ is an {\it almost} one-to-one function, from \cref{eq:congestion_analysis_bound_PI_2}, we can further bound the probability mass flow through $(z,z')$ as
    \begin{align}
        & (W_{\textup{total}})^2 \sum_{\substack{x,y\,:\, \left( x  \in\, \CC_0 \text{ or } \, y \in \CC_0\right),\\ \left(x,y \in \CC_0\cup\CC_2\right) \text{ and }(z,z') \in cp(x,y)}}\pi(x)\pi(y) \nonumber\\
        &\leq~~  W(z) \Bigg(\mathcal{O}(1) \times \sum_{\xi \in \mathcal{C}_0}W(\xi) \nonumber\\
        &\quad\quad+~~ \mathcal{O}(1) \times \frac{2}{(M_1 +2 M_2)} \sum_{\xi \in \mathcal{C}_2}W(\xi)\Bigg)  \\
        &=~~ \mathcal{O}(1)\times W(z)\times W_{\textup{total}}
    \end{align}
    The extra factor of $\mathcal{O}(1)$ is due to the fact that $\eta_{(z,z')}$ is not a perfect one-to-one function.
    Dividing the left and the right hand side by $(W_{\textup{total}})^2$, we get
    \begin{align}
    \label{eq:canonical_path_commodity_bound_1}
       \sum_{\substack{x,y\,:\, \left( x  \in\, \CC_0 \text{ or } \, y \in \CC_0\right),\\ \left(x,y \in \CC_0\cup\CC_2\right) \text{ and }(z,z') \in cp(x,y)}}\pi(x)\pi(y)
       ~\leq~\mathcal{O}(1)\,\pi(z).
    \end{align}
    Since the left hand side of the above inequality is the flow going through the edge $(z,z')$ where $z \in \CC_0$, substituting \cref{eq:canonical_path_commodity_bound_1} into \cref{defeq:congestion}, we get an upper bound on the congestion created by the canonical paths at the edge $(z,z')$ where $z\in \CC_0$ as follows
    \begin{align}
        &\frac{1}{\pi(z)P(z, z')}\sum_{\substack{x,y\,:\, \left( x  \in\, \CC_0 \text{ or } \, y \in \CC_0\right),\\ \left(x,y \in \CC_0\cup\CC_2\right) \text{ and }(z,z') \in cp(x,y)}}\hspace{-.25cm}\pi(x)\pi(y)\\
        &\leq~~ \frac{\mathcal{O}(1)}{P(z, z')}  \\
        &=~~ \left(\min_{i,j}\{a_{ij}-b_{ij}, a_{ij}+b_{ij}\}\right)^{-1} \mathcal{O}(n^6\beta^3\epsilon^{-2})
    \end{align}
    where we have used the following lower bound
    \begin{align}
        P(z,z')
        &\geq \left(\min_{i,j}\{a_{ij}-b_{ij}, a_{ij}+b_{ij}\}\right) {\Omega}\left(L^{-1}\beta\cdot M^{-1}\right)\nonumber\\
        &\hspace{-0.4cm}\geq \left(\min_{i,j}\{a_{ij}-b_{ij}, a_{ij}+b_{ij}\}\right) {\Omega}\left(n^{-6}\beta^{-3}\epsilon^2\right)
    \end{align}
    on the non-zero probability of transition $P(z,z')$ from $z$ to $z'$ when $z \in \CC_0$ (case \#1 of $\mathcal{M}$).

    \item 
    If $z \in \CC_2$ and $x, y \in \CC_0$, then $\eta_{(z,z')}(x,y)\in \CC_2$ and the two additional Pauli-$X$ operators in $\eta_{(z,z')}(x,y)$ will be in the exact same positions as in configuration $z$.  
    Therefore we can bound the sum of weights of configurations $\eta_{(z,z')}(x,y)$ for $z \in \CC_2$, where the summation goes over $x, y \in \CC_0$, as
    \begin{align}
        &\sum_{\substack{x,y\,:\, \left( x  \in\, \CC_0 \text{ and } \, y \in \CC_0\right)\\ \text{ and }(z,z') \in cp(x,y)}} W(\eta_{(z,z')}(x,y))\nonumber\\
        &\leq
        \sum_{\substack{\xi\in\CC_2\\
        \textup{worm heads fixed as in } z}}
        \hspace{-5mm}
        W(\xi) ~~\leq~~~ \mathcal{O}(1) \sum_{\xi \in \CC_0} W(\xi). \label{eq:extended_config_bound_for_C2_encoding_function_PI}
    \end{align}
    The second inequality is equivalent to the bound we evaluate in  
    \cref{ineq:fixedwormC2sum}.
    
    If $z \in \CC_2$ and exactly one of $x, y \in \CC_2$, then $\eta_{(z,z')}(x,y) \in \CC_0 \cup \CC_2 \cup \CC_4$ but whenever $\eta_{(z,z')}(x,y) \in \CC_4$, out of the four additional Pauli-$X$ operators in $\eta_{(z,z')}(x,y)$ two of the Pauli-$X$ operators will be in the exact same positions as in configuration $z$.
    Therefore, similar to \cref{eq:extended_config_bound_for_C2_encoding_function_PI}, we can bound the the sum of weights of configurations $\eta_{(z,z')}(x,y)$ for $z \in \CC_2$, where the summation goes over $x, y \in \CC_0\cup \CC_2$ and exactly one of $x,y \in \CC_2$, as    
    \begin{align}
        \label{eq:extended_config_bound_for_C4_encoding_function_PI}
        \sum_{\substack{x,y\,:\, \text{exactly one of } x,y \in \CC_2,\\ \left(x,y \in \CC_0\cup\CC_2\right) \text{ and }(z,z') \in cp(x,y)\}}} &W(\eta_{(z,z')}(x,y)) \nonumber\\
        &\leq~~ \mathcal{O}(1) \sum_{\xi \in \CC_2} W(\xi).
    \end{align}
    The proof of \cref{eq:extended_config_bound_for_C2_encoding_function_PI} can be easily extended to prove \cref{eq:extended_config_bound_for_C4_encoding_function_PI}. Substituting \cref{eq:extended_config_bound_for_C2_encoding_function_PI} and \cref{eq:extended_config_bound_for_C4_encoding_function_PI} back into \cref{eq:congestion_analysis_bound_PI_2}, we get
    \begin{align}
        &(W_{\textup{total}})^2 \hspace{-0.5cm}\sum_{\substack{x,y\,:\, \left( x  \in\, \CC_0 \text{ or } \, y \in \CC_0\right),\\ \left(x,y \in \CC_0\cup\CC_2\right) \text{ and }(z,z') \in cp(x,y)}}\hspace{-0.5cm}\pi(x)\pi(y) \nonumber\\
        &\leq~~ \mathcal{O}(1) W(z) \left( \sum_{\xi \in \mathcal{C}_0}W(\xi) +  \frac{2}{(M_1 +2 M_2)} \sum_{\xi \in \mathcal{C}_2}W(\xi)\right)
    \end{align}
    Dividing the left and right hand side by $(W_{\textup{total}})^2$, we get
    \begin{align}
        \label{eq:canonical_path_commodity_bound_2}
        \sum_{\substack{x,y\,:\, \left( x  \in\, \CC_0 \text{ or } \, y \in \CC_0\right),\\ \left(x,y \in \CC_0\cup\CC_2\right) \text{ and }(z,z') \in cp(x,y)}}\hspace{-0.5cm}\pi(x)\pi(y) \leq \mathcal{O}(M_1+2M_2)\,\pi(z)
    \end{align}
    Since the left hand side of the above inequality is the flow going through the edge $(z,z')$ where $z \in \CC_2$, substituting \cref{eq:canonical_path_commodity_bound_2} into \cref{defeq:congestion}, we get an upper bound on the congestion created by the canonical paths at the edge $(z,z')$ where $z\in \CC_2$ as follows
    \begin{align}
        &\frac{1}{\pi(z) P(z,z')}\hspace{-0.5cm}\sum_{\substack{x,y\,:\, \left( x  \in\, \CC_0 \text{ or } \, y \in \CC_0\right),\\ \left(x,y \in \CC_0\cup\CC_2\right) \text{ and }(z,z') \in cp(x,y)}}\hspace{-0.5cm}\pi(x)\pi(y) \nonumber\\
        &\leq~~ \frac{\mathcal{O}(M_1+2M_2)}{P(z,z')}\\
        &= \left(\min_{i,j}\{a_{ij}-b_{ij}, a_{ij}+b_{ij}\}\right)^{-1} \mathcal{O}(n^6\beta^3\epsilon^{-2})
    \end{align}
    where we have used $M = M_1+2M_2 = \mathcal{O}(n^4\beta^{2}\epsilon^{-1})$ and the following lower bound
    \begin{align}
        P(z,z')
        \geq~~ \left(\min_{i,j}\{a_{ij}-b_{ij}, a_{ij}+b_{ij}\}\right) {\Omega}\left(L^{-1}\beta\right) \nonumber\\
        \geq~~ \left(\min_{i,j}\{a_{ij}-b_{ij}, a_{ij}+b_{ij}\}\right) {\Omega}\left(n^{-2}\beta^{-1}\epsilon\right)
    \end{align}
    on the non-zero probability of transition $P(z,z')$ form $z$ to $z'$ when $z \in \CC_2$. 
\end{enumerate}

So far, we have constructed the flow of probability mass between any two configurations where at least one of the configurations is in $\mathcal{C}_0$. We are left to construct the flow between configurations in $\mathcal{C}_2$. Our strategy is to distribute the probability mass between any two configurations in $\CC_2$ through the configurations in $\CC_0$. Specifically, consider two configurations $x,y \in \CC_2$. We will first route $\pi(x)\pi(y)\pi(w)/\sum_{\xi\in \CC_0} \pi(\xi)$ amount of probability mass from $x$ to a configuration $w \in \CC_0$ via the canonical path we constructed previously from $x$ to $w$ and route the same amount from $w$ to $y$ via the canonical path from $w$ to $y$. Summing over the configurations in $\CC_0$ will allow us to route $\pi(x)\pi(y)$ amount of probability mass from $x$ to $y$. We can bound the total amount of extra congestion created at $(z,z')$ from these additional flows as
\begin{widetext}
\begin{align}
    &\frac{1}{\pi(z)P(z,z')}\sum_{\substack{x,y,w\,:\, x,y \in \CC_2\text{ and }w \in \CC_0\\ \text{ and }(z,z') \in cp(x,w)\text{ or } cp(w,y)}} \frac{\pi(x)\pi(y)\pi(w)}{\sum_{\xi\in \CC_0} \pi(\xi)}\nonumber\\
    &=~~ \frac{1}{\pi(z)P(z,z')}\cdot\frac{\sum_{y \in \CC_2}\, \pi(y) }{\sum_{\xi\in \CC_0} \pi(\xi)}\left(\sum_{\substack{x,w\,:\, x\in \CC_2, w\in\CC_0\\ \text{ and } (z,z') \in cp(x,w) \text{ or }cp(w,x)}} \hspace{-0.25cm}\pi(x) \pi(w)\right)\\
    &=~~ \frac{\sum_{y \in \CC_2}\, \pi(y) }{\sum_{\xi\in \CC_0} \pi(\xi)}\left(\frac{1}{\pi(z)P(z,z')}\sum_{\substack{x,w\,:\, x\in \CC_2, w\in\CC_0\\ \text{ and } (z,z') \in cp(x,w) \text{ or }cp(w,x)}}\hspace{-0.25cm} \pi(x) \pi(w)\right)\\
    &\leq~~ \frac{\sum_{y \in \CC_2}\, \pi(y) }{\sum_{\xi\in \CC_0}\pi(\xi)}\cdot \left(\min_{i,j}\{a_{ij}-b_{ij}, a_{ij}+b_{ij}\}\right)^{-1} \mathcal{O}(n^6\beta^3\epsilon^{-2}) \\
    &\leq~~ \left(\min_{i,j}\{a_{ij}-b_{ij}, a_{ij}+b_{ij}\}\right)^{-1} \mathcal{O}(n^{10}\beta^{5}\epsilon^{-3})
\end{align}
\end{widetext}
where the first inequality is due to \cref{eq:canonical_path_commodity_bound_1} and \cref{eq:canonical_path_commodity_bound_2}, and the second inequality is due to \cref{thm:polynomial_bound_on_extended_config_weight_PI}. 

Considering the configuration $\sigma_m = \ket{0}^{\otimes n}$ for all $0\leq m \leq M$ as the initial state $\gamma_{\textup{initial, PI}}$ to the Markov chain $\mathcal{M}$ where $\ln(\pi(\gamma_{\textup{initial, PI}})) \leq \mathcal{O}(1)\beta||H||_{\textup{op}} \leq \mathcal{O}(1) n^2\beta$, by \cref{thm:mixing_time_theorem}, we can bound the mixing time of the Markov chain $\mathcal{M}$ as
\begin{align}
    t_{mix} ~&=~ c_{\min}^{-1}\,\tilde{\mathcal{O}}(n^{10}\beta^{5}\epsilon^{-3})\cdot M\cdot n^2\beta  \\
    &=~ c_{\min}^{-1}\,\tilde{\mathcal{O}}(n^{16}\beta^{8}\epsilon^{-4})
\end{align}
where $c_{\min} = \left(\min_{i,j}\{a_{ij}-b_{ij}, a_{ij}+b_{ij}\}\right)$.

\subsubsection{Constructing of encoding function $\eta$}
\label{sec:encoding_function_construction}
Let us now define the encoding  function $\eta_{(z,z')}$ which encodes the canonical path from $x$ to $y$ that goes through the edge $(z,z')$ and satisfies the weight condition in  \cref{eq:encoding_function_condition_PI}.  Since the weights of the configurations are defined as a product of individual operator weights,  we will define encoding configuration by defining the legs of each operator such that operator weights satisfy 
\begin{align}
    \label{eq:encoding_function_operator_weight_condition}
    &\langle\sigma_m(x)|O_m|\sigma_{m+1}(x)\rangle\times \braket{\sigma_m(y)|O_m|\sigma_{m+1}(y)}\nonumber\\
    &\hspace{0.5cm}\leq~~ \braket{\sigma_m(z)|O_m|\sigma_{m+1}(z)}\nonumber\\
    &\hspace{1.5cm} \times \braket{\sigma_m(\eta_z(x,y))|O_m|\sigma_{m+1}(\eta_z(x,y))}
\end{align}
for every operator $O_m$.

\begin{figure*}[!th]
    \centering
    \begin{subfigure}[c]{0.45\textwidth}
        \centering
        \includegraphics[scale=0.75]{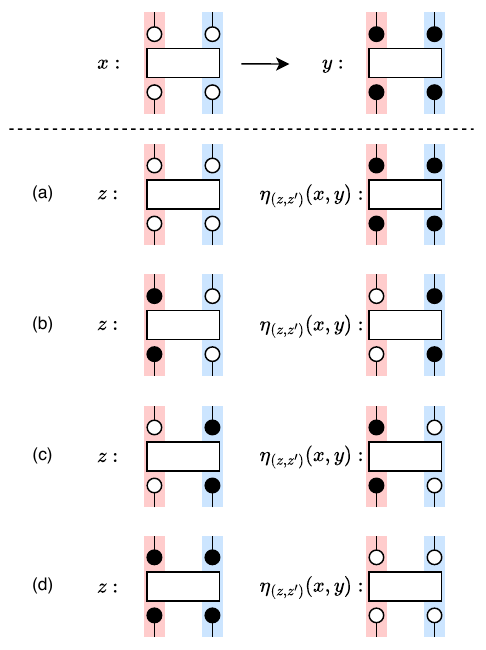}
        \caption{}
        \label{fig:encoding_func_example1}
    \end{subfigure}
    \begin{subfigure}[c]{0.45\textwidth}
        \centering
        \includegraphics[scale=0.75]{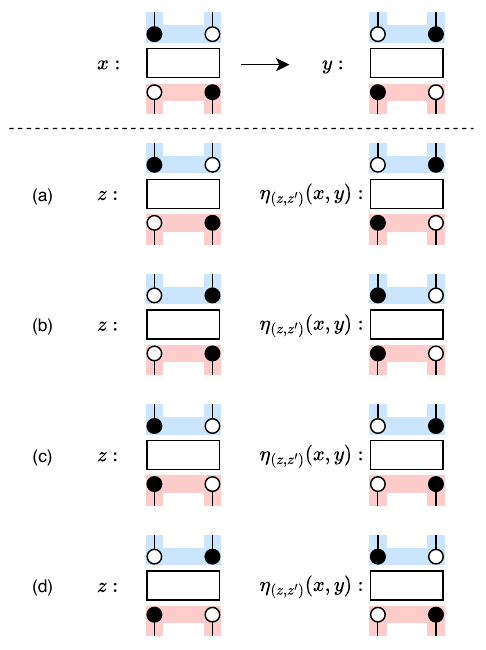}
        \caption{}
        \label{fig:encoding_func_example2}
    \end{subfigure}
    \begin{subfigure}[c]{0.45\textwidth}
        \centering
        \includegraphics[scale=0.75]{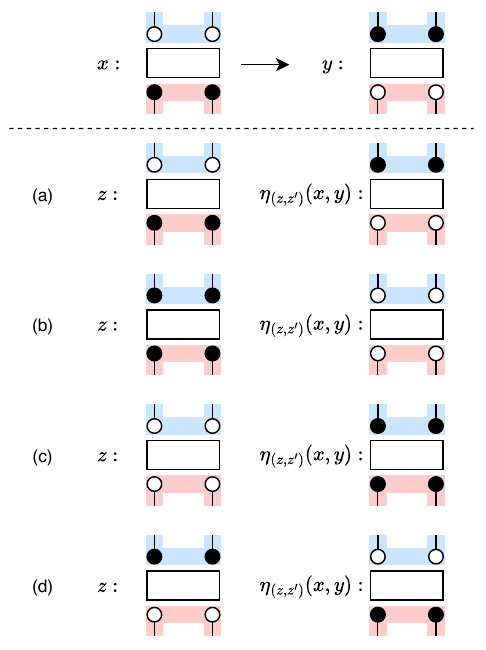}
        \caption{}
        \label{fig:encoding_func_example3}
    \end{subfigure}
    \begin{subfigure}[c]{0.45\textwidth}
        \centering
        \includegraphics[scale=0.75]{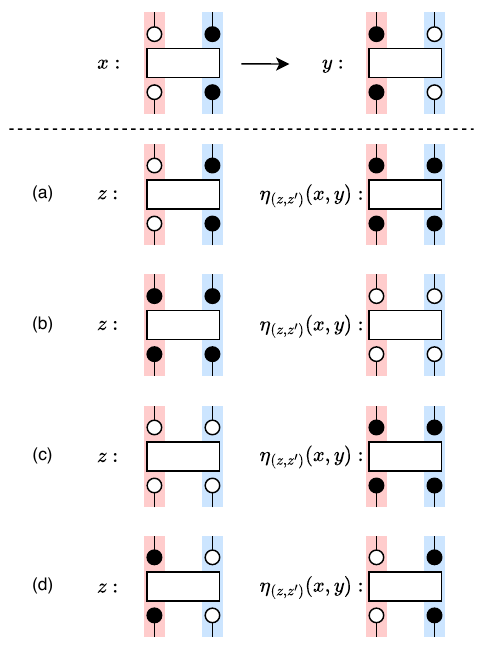}
        \caption{}
        \label{fig:encoding_func_example4}
    \end{subfigure}
    \caption{
    List of the encoding function $\eta_{(z,z')}(x,y)$ configuration at an operator $O_m$ for different cases of the $z$ configuration along the canonical path between $x$ and $y$ where all four legs of $O_m$ are different. 
    }
    \label{fig:encoding_func_examples}
\end{figure*}

\begin{enumerate}[leftmargin=*]
\item Consider an operator $O_m$ that has two of its legs different between the configurations $x$ and $y$. Since the canonical path is construct by unwinding  the loops and the open path, the configuration $z$ along the canonical path between $x$ and $y$ will have the legs of the operator $O_m$ matching either $x$ or $y$. If the legs of the operator $O_m$ in the $z$ configuration match the $x$ configuration, we define the encoding configuration with the legs of the operator $O_m$ matching the $y$ configuration and vice versa. In this case, it is easy see that the weights coming from the operator $O_m$ satisfies the inequality \cref{eq:encoding_function_operator_weight_condition}.

\item If an operator $O_m$ has four of its legs different between the configurations $x$ and $y$, then the configuration $z$ along the canonical path can either have all the four legs of the operator $O_m$ match the configuration $x$ or $y$ or have two of its legs match $x$ and the other two legs match $y$. In these cases, we define the encoding configuration with all the four legs of the operator $O_m$  flipped from the configuration $z$. \Cref{fig:encoding_func_examples} gives four examples of an operator in $x$ and $y$ where all the four legs of an operator are different and the possible $z$ configurations along the canonical path and the corresponding encoding configurations. Based on the convention we fixed on how to decompose loops as shown in \cref{fig:loop_decomposition}, we can verify that the weights coming from the operator $O_m$ satisfy the inequality \cref{eq:encoding_function_operator_weight_condition}.
\end{enumerate}

\begin{figure*}[!th]
    \centering
    \begin{subfigure}[c]{0.45\textwidth}
        \centering
        \includegraphics[scale=0.75]{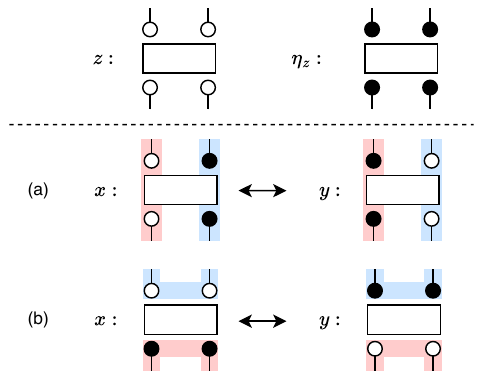}
        \caption{}
        \label{fig:decoding_encoding_func_example1}
    \end{subfigure}
    \hfill
    \begin{subfigure}[c]{0.45\textwidth}
        \centering
        \includegraphics[scale=0.75]{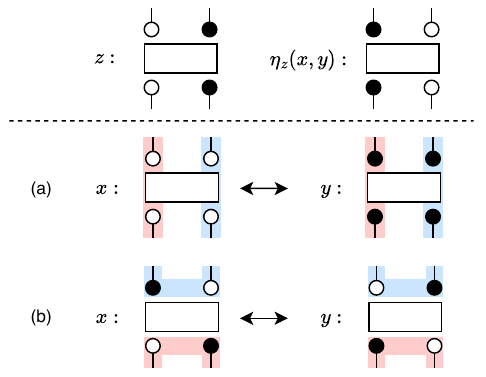}
        \caption{}
        \label{fig:decoding_encoding_func_example2}
    \end{subfigure}
    \caption{
    Potential ambiguities for the $x$ and $y$ configurations when $z$ and $\eta_{(z,z^{\prime})}$ configurations have different states for all four legs at a particular operator. 
    }
    \label{fig:decoding_encoding_func_examples}
\end{figure*}

A couple of observations about the encoding function are in order. First, we defined the encoding function by defining the legs of each of the operators $O_m$ but one can check that legs of the neighboring operators are consistent with each other except at a maximum of four places where we need to add the additional Pauli-$X$ terms for consistency in the diagrammatic notation. Therefore,  $\eta_{(z,z')} (x,y)$ is always in $\CC_0 \cup \CC_2 \cup \CC_4$. Specifically, 
\begin{enumerate}[leftmargin=*]
    \item If $x,y,z \in \CC_0$, the encoding function $\eta_{(z,z')} (x,y)$ defined above doesn't need any additional Pauli-$X$ operators for consistency in the imaginary time direction and therefore $\eta_{(z,z')} (x,y) \in \CC_0$. 
    \item If $x,y \in \CC_0$ and $z \in \CC_2$, then $\eta_{(z,z')} (x,y)$ will need two additional Pauli-$X$ operators in the same place as in $z$ for consistency in the imaginary time direction and therefore $\eta_{(z,z')} (x,y) \in \CC_2$.
    \item If $x \in \CC_0$ \& $y \in \CC_2$ or $x \in \CC_2$ \& $y \in \CC_0$ and $z \in \CC_0$, then $\eta_{(z,z')} (x,y)$ will need two additional Pauli-$X$ operators in the same place as the two additional Pauli-$X$ operators in $x$ or $y$, whichever is in $\CC_2$, for consistency in the imaginary time direction and therefore $\eta_{(z,z')} (x,y) \in \CC_2$.
    \item If $x \in \CC_0$ \& $y \in \CC_2$ or $x \in \CC_2$ \& $y \in \CC_0$ and $z \in \CC_2$, then $\eta_{(z,z')} (x,y)$ will need zero or two or four additional Pauli-$X$ operators for consistency in the imaginary direction. When $\eta_{(z,z')} (x,y) \in \CC_4$, two of the four additional Pauli-$X$ operators have to be in the same place as the two additional Pauli-$X$ operators in $z$.
\end{enumerate} 

Second, we claim that the encoding function is $\eta_{(z,z')}$ is {\it almost} one-to-one. Suppose that we have a two-qubit operator $O_m$ with its legs in state $\bra{00}$ and $\ket{00}$ in the configuration $z$ and the same operator has its legs in state $\bra{11}$ and $\ket{11}$ in the configuration $\eta_{(z,z')}(x,y)$ or $\bra{01}$ and $\ket{01}$ in $z$ and $\bra{10}$ and $\ket{10}$ in $\eta_{(z,z')}(x,y)$. This implies that all the four legs of the operator $O_m$ are different between $x$ and $y$. When $x$ and $y$ have an operator $O_m$ with all four its legs different between them then depending on the state of the legs of $O_m$ in $x$ and $y$, we have a fixed convention to decompose the loops and paths either horizontally or vertically at that particular operator $O_m$, as shown in \cref{fig:loop_decomposition}. Once the horizontal or vertical decomposition is decided for every operator $O_m$ that has four of its legs different between $z$ and $\eta_{(z,z')}(x,y)$, and therefore between $x$ and $y$, we can use the canonical order of the unwinding to uniquely deduce the configurations $x$ and $y$ given $z$ and $\eta_{(z,z')}$. This ambiguity, as shown in \cref{fig:decoding_encoding_func_examples}, in deciding whether to decompose the loops and path at an operator $O_m$ when it has all four or its legs different between $z$ and $\eta_{(z,z')}$ is the reason why $\eta_{(z,z')}$ is not a one-to-one encoding function.

Nevertheless, we can still bound the sum of product of weights $W(x)W(y)$ of all possible configurations $x$ and $y$ given a fixed $z$ and $\eta_{(z,z')}$ and upper bound it with $W(z)W(\eta_{(z,z')})$ required as \cref{eq:encoding_function_operator_weight_condition}. 
When the loop at the operator $O_m$ is decomposed horizontally in $x$ and $y$, the product of operator weights of $O_m$ in $x$ and $y$ is $\leq \delta^2 ={\beta^2}/{4L^2}$. 
On the other hand, when the loop at the operator $O_m$ is decomposed vertically in $x$ and $y$, the product of operator weights of $O_m$ in $x$ and $y$ is equal to 1. Considering these two possibilities, we get $(1+{\beta^2}/{4L^2})$ for each such operator $O_m$ and taking a product for all such operator weights in $x$ and $y$ which there could be a maximum of $M = \mathcal{O}(n^2L)$, we get
\begin{align}
    \left(1+\frac{\beta^2}{4L^2}\right)^{M} ~~\leq~~ e^{\mathcal{O}(1)\beta^2L^{-1}n^2} ~~\leq~~ \mathcal{O}(1).
\end{align}
Therefore, even though $\eta_{(z,z')}$ is not a one-to-one encoding function, we can treat it as such for the purpose of bounding the probability mass flow in the path-integral congestion analysis with a constant overhead.

\section{Discussions}
\label{sec:discussion}

A fast-mixing proof of QMC for a family of Hamiltonians does not only establish algorithmic efficiency, but also carries implications of interest in condensed matter physics.
Here, we outline some of the most intriguing directions along these lines.
Turning these observations into formal consequences, i.e., deriving rigorous implications from our theorems, must be quite nontrivial. 
However, we believe that the physical arguments we provide here offer sufficiently clear conceptual bridges that may lead to future mathematical formulations. 
In addition, we highlight several specific open problems directly related to our work in the final subsection.

\subsection{Implications and Relations to Condensed-Matter and Many-Body Physics}

\begin{enumerate}[leftmargin=*]

    \item The fact that the operator-loop update QMC can mix in polynomial time prohibits some phases of matter and also phase transitions to occur as the thermal/ground state of this class of Hamiltonians. 

    The first of such kind is a discrete spontaneous symmetry (SSB) breaking phase, characterized by an order parameter expressible in the $Z$-basis. 
    This includes, for example, valence-bond solid phases in which the discrete lattice symmetries are spontaneously broken. This phase was studied in, e.g., \cite{Ma18DynamicalSignatureDQC,Zhao25ScalingOperatorEPDQC} where an extended version of the $XY$ model with four-body interactions was used to realize the valence-bond solid phase. 
    Our claim is that such an extension of the Hamiltonian to include four-body interactions is {\it necessary} as long as one wishes to remain within the framework of stoquastic $XY$ models. 
    The reason is straightforward: when the thermal/ground state has a discrete SSB, any QMC that only changes the configuration locally at a step --- this includes the operator-loop update QMC we consider --- must take an exponential time to mix. Since we exclude that possibility, the conclusion is that discrete SSB cannot arise in this class of Hamiltonians. 
    The intuition is that in order to move from one discrete SSB sector to another (e.g. from the all-up state to all-down for an Ising model), one must go through a configuration with a domain wall that grows as $\mathcal O(L^{d-1})$ in size, with higher energy and exponentially smaller probabilities. 
    In some cases, this type of exponentially slow mixing time is rigorously proven \cite{cesi1996_2dIsingSlowInAllLowTemp, schonmann1994slowdroplet}. 
    
    On the other hand, continuous symmetry SSB states are known to be polynomial mixing, due to the existence of Goldstone modes \cite{becker2020spectral,Chaikin_Lubensky_1995}. 
    In essence, continuous symmetries permit infinitesimal deformations of the order parameter, avoiding the energetic cost of sharp domain walls.
    This is consistent with the fact that the $XY$ model on the square lattice is known to host $U(1)$ SSB phases as the ground state \cite{Kennedy88XYhasLRO}. 

    We note that the transverse-field Ising model obtained by setting $b_{ij}=0$ does host a discrete $\mathbb{Z}_2$ SSB phase. However, the order resides in the $X$-direction and is thus invisible to any order parameter defined in the $Z$-basis.
    Unless the $Z$-basis configuration (which is the one we use in the QMC we consider) can tell which symmetry sector the configuration is in, the above logic does not work. 

    Another phenomenon that is prohibited by our result is a first-order phase transition that accompanies phase coexistence at the transition point. 
    The reasoning is almost exactly the same as the one for discrete SSB, except that here no assumption about a $Z$-basis order parameter is needed.
    This is because at first-order transitions thermodynamic quantities change discontinuously, and this can be seen with a difference in the $Z$-operator appearance frequency etc. in the configuration. 
    This also implies that with any variant of the transverse field Ising model, as long as it only has ferromagnetic two-body interaction, it can never host a standard first-order transition 
    \footnote{It is very intriguing that there actually {\it exists} a first-order transition in a ferromagnetic Ising model but barely evades contradiction by not having a phase coexistence. This is the peculiar case of a 1-dimensional Ising model with long-range interaction that decays as $1/r^2$ \cite{aizenman2015IsingContinuity,Kosterlitz2016KTreview}. In this case, the magnetization is actually mappable to the spin stiffness of the standard Kosterlitz-Thouless transition setup in 2d, and therefore has a jump. However, due to the weakness of the KT transition, this transition does not exhibit phase coexistence at the transition point --- a fact that in some sense, is an implication of our result or other general fast-mixing results for the ferromagnetic Ising model \cite{guo2017random}.}.

    \item 
    A more quantitative implication of our result is that it provides a rigorous upper bound on the algorithmic dynamical exponent $z_{\mathrm{alg}}$ for the operator-loop update QMC applied to various stoquastic $XY$ models.
    The exponent $z_{\mathrm{alg}}$ characterizes the dynamical slowdown of the QMC algorithm and is defined by the scaling $t_{\mathrm{mix}}(n) \sim n^{z_{\mathrm{alg}} + 1}$ with system size $n$. The additional $+1$ accounts for the fact that a single {\it Monte Carlo step} consists of $\mathcal{O}(n)$ operator-loop QMC steps in our convention.

    Our result $t_{\mathrm{mix}} \lesssim \tilde{\mathcal{O}}(n^{8})$ for lattice systems translates to an upper bound $z_{\mathrm{alg}} \leq 7$, which remains quite loose compared to the empirically observed value $z_{\mathrm{alg}} \sim 2$ reported in \cite{ZYdynamical} for closely related systems.
    Nevertheless, rigorous results on $z_{\mathrm{alg}}$ for quantum systems are scarce, and we believe that even a loose upper bound provides a valuable starting point—particularly in light of recent interest in dynamical exponents from the perspective of quantum information theory and field-theoretic approaches \cite{masaoka2025rigorous, isakov2011dynamics}.

    \item 
    Quantum Gibbs state preparation and sampling has recently attracted significant attention as a promising avenue for demonstrating quantum advantage \cite{Poulin09PRL,zhan2025rapid,ding2025efficient,lin2025dissipative}.
    While numerous quantum protocols have been proposed, relatively little was known about how much of this task is classically achievable in polynomial time.
    This gap becomes particularly evident when we note that many benchmark problems used in these protocols involve quantum spin models—such as the transverse-field Ising, Heisenberg, or $XY$ models.
    A crucial point is that these models are stoquastic and thus amenable to QMC. In particular, we prove that for stoquastic XY models, thermal sampling can always be performed in polynomial time using QMC, a purely classical algorithm.

    Of course, QMC sampling and quantum Gibbs state preparation differ in capability. The latter yields an actual quantum state, allowing measurement of arbitrary observables. In contrast, QMC sampling gives direct access to $Z$-basis observables and Hamiltonian expectation values, but evaluating general observables can become practically challenging.
    One may argue that a perturbatively small external field aligned with the desired observable could allow indirect estimation via QMC. It remains an interesting open question whether this procedure incurs only polynomial overhead, and more generally, what the essential computational differences are between QMC sampling and quantum Gibbs state preparation \cite{yin2023polynomialtimeclassicalsamplinghightemperature}.

\end{enumerate}

\subsection{Open problems}
\begin{enumerate}[leftmargin=*]
    \item A significant open problem, the original motivation of our work, is to extend our result to show that the operator-loop QMC is fast mixing for stoquastic Heisenberg Hamiltonians\label{sec:HeisenbergOpenProblem}
    \begin{align}
        H = \sum_{ij} w_{ij}(-X_iX_j+Y_iY_j-\Delta Z_iZ_j)
    \end{align}
    in the easy-plane region $-1 \leq \Delta \leq 1$ which includes quantum max-cut on bipartite graphs after a local basis transformation. Empirically, it is well-known that the operator-loop QMC is fast mixing on common bipartite lattices with Heisenberg interaction. 
    The complexity of quantum max-cut on bipartite lattices is explicitly raised as an open problem in \cite{gharibian7faces}. 
    A bound on the mixing time of the operator-loop QMC for this class of Hamiltonians will not only resolve the open problem but also fill the gap between our current mathematical understanding and the empirical physical observations. 

    \item The operator-loop update with path-integral QMC was one of the first algorithms that has been successfully used in practice, but there have been other QMC methods \cite{syljuaasen2002quantum, evertz2003loop} that are more efficient in practice. 
    One such algorithm called the directed-loop QMC by \cite{syljuaasen2002quantum} is a lifted version of the operator-loop QMC we considered here, where it breaks the detailed balance condition to sample from the QMC configuration space more efficiently. 
    Intuitively, rather than letting the worms of the operator-loop wander around as in a random walk (diffusive, $\sim\mathcal O(\sqrt t)$), it is better to have them move in a biased direction (ballistic, $\sim\mathcal O(t)$). 
    In many cases, irreversible Markov chains which break the detailed balance condition mix faster than their reversible counterparts \cite{chen1999lifting,  nishikawa2015event}.  Therefore, it is an interesting open direction to bound the runtime of the directed-loop QMC to get a more efficient algorithm for the stoquastic XY Hamiltonians.

    \item While QMC algorithms and classical Markov chains for sampling from thermal distributions of classical Hamiltonians are usually different in their inceptions and therefore different in their character, when there is a connection that can be drawn between the two extremes, it has the potential to give useful insights on both ends. Evertz's loop algorithm \cite{evertz2003loop} is one such connection. The loop algorithm generalizes both the Swendsen-Wang algorithm for classical Ferromagnetic Ising Hamiltonians as well the operator-loop QMC for Heisenberg and XY Hamiltonians. Empirically, the loop algorithm seems to be efficient in all the regimes interpolating between the stoquastic XY Hamiltonian and the Ferromagnetic Ising Hamiltonian. 
    However, it is only recently when the Swendsen-Wang algorithm is {\it proven} to be efficient \cite{guo2017random}, while the efficiency of the QMC end of the loop algorithm is still to be proven. We conjecture that the loop algorithm is efficient all the way from the Swendsen-Wang algorithm on the classical end to the operator-loop QMC for the stoquastic XY Hamiltonians.
\end{enumerate}

\acknowledgements
We thank Elizabeth Crosson, Samuel Slezak, Ojas Parekh, Andrew Zhao, Naoki Kawashima, and Anders Sandvik for valuable discussions. This work was supported by the U.S. Department of Energy, Office of Science, Accelerated Research in Quantum Computing, Fundamental Algorithmic Research toward Quantum Utility (FAR-Qu) (CR), 
and the Japanese Ministry of Education, Culture, Sports, Science and Technology (MEXT) KAKENHI Grant numbers JP25034663, JP25034741, and JP25099514 (JT). 
The authors also thank the QIP 2025 conference, where this research was first conceived.

\bibliographystyle{alpha}
\bibliography{refs}

\onecolumngrid
\appendix

\section{Matrix form of a path-integral operator in computational basis}

The operator $I+\delta \left(a_{ij} X_iX_j -b_{ij}Y_iY_j\right)$ has the following explicit form in the computational basis.
\begin{align}
    \label{eq:two_operator_matrix_form}
    I+\delta \left(a_{ij} X_iX_j -b_{ij}Y_iY_j\right) &= 
    \begin{pmatrix}
        1 & 0 & 0 & \delta(a_{ij}+b_{ij})\\
        0 & 1 & \delta(a_{ij}-b_{ij}) & 0\\
        0 & \delta(a_{ij}-b_{ij}) & 1 & 0\\
        \delta(a_{ij}+b_{ij}) & 0 & 0 & 1
    \end{pmatrix}
\end{align}

\section{Counting-to-Sampling reduction}
\label{sec:counting_to_sampling_reduction}
The problem of estimating the Trotterized partition function $\tilde{Z}(\beta, H)$, and therefore the partition function $Z(\beta, H)$ by \cref{thm:trotterization}, can be reduced to the problem of sampling from the distribution $\pi_{0}$, using a counting-to-sampling argument \cite{jerrum1993polynomial, crosson2021rapid} as follows. Let us rewrite $\tilde{Z}(\beta, H)$ as a telescoping product
\begin{align}
    \label{eq:telescoping_product}
    \tilde{Z}(\beta, H) = \tilde{Z}(\beta_0, H) \prod_{i=1}^{k}\frac{\tilde{Z}(\beta_i, H)}{\tilde{Z}(\beta_{i-1}, H)}
\end{align}
where $\beta_0 = 0$ and $\beta_k = \beta$. Each of the fractions $\tilde{Z}(\beta_i, H)/\tilde{Z}(\beta_{i-1}, H)$ in the product can be estimated separately as an observable
\begin{align}
    \left\langle \frac{W_{\beta_{i}}}{W_{\beta_{i-1}}}\right\rangle_{\pi_{\beta_{i-1}}} &= \sum_{\sigma_1,...\sigma_M} \frac{W_{\beta_{i}}(\sigma_1,...,\sigma_M)}{W_{\beta_{i-1}}(\sigma_1,...,\sigma_M)} \pi_{\beta_{i-1}}(\sigma_1,...,\sigma_M)\\
    &= \sum_{\sigma_1,...\sigma_M} \frac{W_{\beta_{i}}(\sigma_1,...,\sigma_M)}{W_{\beta_{i-1}}(\sigma_1,...,\sigma_M)} \frac{W_{\beta_{i-1}}(\sigma_1,...,\sigma_M)}{\tilde{Z}(\beta_{i-1}, H)} \\
    &= \frac{\sum_{\sigma_1,...\sigma_M} W_{\beta_{i}}(\sigma_1,...,\sigma_M)}{\tilde{Z}(\beta_{i-1}, H)} = \frac{\tilde{Z}(\beta_i, H)}{\tilde{Z}(\beta_{i-1}, H)} \label{eq:ratio_of_partition_functions}
\end{align}
if we have access to the samples from the distribution $\pi_{0,\beta_{i-1}}$. The subscripts for $\pi_0$ and $W$ indicate their explicit dependence on the inverse temperature $\beta$ for a fixed number of Trotter steps $L$. Choosing $\beta_i-\beta_{i-1} = 1/||H||$ ensures that $\tilde{Z}(\beta_i, H)/\tilde{Z}(\beta_{i-1}, H) \approx Z(\beta_i, H)/Z(\beta_{i-1}, H) \in [e^{-1},e]$. To estimate $\tilde{Z}(\beta, H)$ to within a $\exp({\pm \epsilon})$ multiplicative precision, we want to estimate the ratio in \cref{eq:ratio_of_partition_functions} to within a $\exp({\pm \epsilon/\beta||H||_{\textup{op}}})$ precision since the telescoping product in \cref{eq:telescoping_product} has $\beta||H||_{\textup{op}}$ number of terms when we choose $\beta_i-\beta_{i-1} = 1/||H||$. This implies, by Hoeffding's inequality, we will need to take $\tilde{\mathcal{O}}(\beta^2||H||_{\textup{op}}^2\epsilon^{-2}) = \tilde{\mathcal{O}}(n^4\beta^2\epsilon^{-2})$ number of samples from the distribution $\pi_{\beta_{i-1}}$ to estimate the ratio in \cref{eq:ratio_of_partition_functions} to within a $\exp({\pm \epsilon/\beta||H||_{\textup{op}}})$ precision with high probability. In total, we need 
\begin{align}
\label{eq:total_number_of_sample_needed}
\tilde{\mathcal{O}}(\beta^3||H||_{\textup{op}}^3\epsilon^{-2}) = \tilde{\mathcal{O}}(n^6\beta^3\epsilon^{-2})
\end{align}
many samples to estimate the Trotterrized partition function $\tilde{Z}(\beta, H)$.

\section{Proof of \cref{thm:polynomial_bound_on_extended_config_weight_PI}}
\label{appendix:proof_of_polynomial_bound_on_extended_config_weight_PI}

Starting from \cref{thm:trotterization} with $L = \mathcal{O}(n^2\beta^{2}\epsilon^{-1})$, we have $\tilde{Z} = \tr\left[(C C^\dagger)^L\right] = \tr\left[\prod_{m=1}^{M} O_m\right]$ which is equal to the sum of weights of all the configurations in $\CC_0$. The operator $C$ is defined in \cref{eq:trotter_expansion} as a product of $\mathcal{O}(n^2)$ operators and each such operator $O_m$ is of the form $\left(I+\delta \left(a_{ij} X_iX_j -b_{ij}Y_iY_j\right)\right)$ or $\left(I- \delta\, d_i Z_i\right)$ with $\delta = \beta/2L$. We use an observation from \cite{bravyi2017polynomial} that the sum of weights of configurations in $\CC_2$ is bounded as
\begin{align}
    \sum_{\xi \in \CC_2} W(\xi) &\leq \mathcal{O}(1)\sum_{1 \leq m_1<m_2\leq M} \tr\left[O_1O_2...O_{m_1} \,X_a\, O_{m_1+1}...O_{m_2}\,X_b\, O_{m_2+1}...O_{M-1}O_M\right]\label{eq:AB_bound1}\\
    &= \mathcal{O}(1)
    \sum_{1 \leq m_1<m_2\leq M} \tr\left[ B(m_1,m_2) X_a A(m_1,m_2) X_b\right] \label{eq:AB_bound2}
\end{align}
where $A(m_1,m_2) = O_{m_1+1}O_{m_1+2}...O_{m_2}$, $B(m_1,m_2) = O_{m_2+1}...O_{M-1}O_MO_1O_2...O_{m_1}$, and $X_a$ and $X_b$ are the two additional Pauli-$X$ operators inserted in between the operator sequence $\{O_m\}_{m=1}^M$ for $\CC_2$ configurations. (see the definition of $\CC_2$ configurations in the path-integral formalism in \cref{sec:operator_update_markov_chain_PI} ). We can bound a single summand from \cref{eq:AB_bound2} which is of the form $\tr\left[ B X_a A X_b\right]$ as follows
\begin{align}
    \label{eq:AB_bound3}
    \tr\left[ B X_a A X_b\right] \leq \sum_k s_k(A)s_k(B)
\end{align}
from von Neumann's trace inequality. 
We denote by $s_k(X)$ the $k^{\textup{th}}$ largest singular value of matrix $X$. 
Observing that the bulk of $A$ and $B$ are just $CC^{\dagger}$ and $||CC^\dagger||_{\textup{op}}\leq2$, we can express them as 
\begin{align}
    \label{eq:A_B_form}
    A = S(CC^\dagger)^{K_a}T \quad \text{and} \quad B = S'(CC^\dagger)^{K_b}T'
\end{align}
where $K_a, K_b \geq 0$ are integers and $||S||_{\textup{op}},\,||T||_{\textup{op}},\,||S'||_{\textup{op}},\,||T'||_{\textup{op}} \leq 2$, 
similarly to eq. (54) of \cite{bravyi2017polynomial}. 
Furthermore, although not explicitly stated in \cite{bravyi2017polynomial}, we can also infer that 
\begin{align}
    L-2 \leq K_a + K_b \leq L
\end{align}
by construction. We can upper bound \cref{eq:A_B_form} with
\begin{align}
    s_k(A) \leq ||S||_{\textup{op}}\cdot||T||_{\textup{op}} \cdot s_k\left((CC^\dagger)^{K_a}\right) \leq \mathcal{O}(1) \left(s_k\left(CC^\dagger\right)\right)^{K_a}
\end{align}
and similarly $s_k(B) \leq \mathcal{O}(1) \left(s_k\left(CC^\dagger\right)\right)^{K_b}$. Substituting these back into \cref{eq:AB_bound3}, we get
\begin{align}
    \tr\left[ B X_a A X_b\right] &\leq \mathcal{O}(1) \sum_k \left(s_k\left(CC^\dagger\right)\right)^{K_a+K_b} \\
    &\leq \mathcal{O}(1) \sum_k \left(s_k\left(CC^\dagger\right)\right)^{L} \\
    &\leq \mathcal{O}(1)\,\tr\left[(CC^\dagger)^L\right] = \mathcal{O}(1) \sum_{\xi \in \CC_0} W(\xi)\label{ineq:fixedwormC2sum}
\end{align}
Substituting the above inequality into \cref{eq:AB_bound2}, we get
\begin{align}
     \sum_{\xi \in \CC_2} W(\xi) \leq \mathcal{O}(1)M^2 \sum_{\xi \in \CC_0} W(\xi) \leq \mathcal{O}((M_1+2M_2)^2)\sum_{\xi \in \CC_0} W(\xi)
\end{align}
Finally, we can bound the ratio of probability of configurations in $\CC_2$ to probability of configurations in $\CC_0$ in the distribution $\pi$ as 
\begin{align}
\label{eq:polynomial_bound_on_extended_config_weight_PI}
    \frac{\sum_{\xi \in \mathcal{C}_2} \pi(\xi)}{\sum_{\xi \in \mathcal{C}_0} \pi(\xi)} = \frac{2}{(M_1 +2 M_2)}\frac{\sum_{\xi \in \mathcal{C}_2} W(\xi)}{\sum_{\xi \in \mathcal{C}_0} W(\xi)} \leq \mathcal{O}(M_1+2M_2) \leq \mathcal{O}(M)\leq \mathcal{O}(n^4\beta^{2}\epsilon^{-1}).
\end{align}

\end{document}